\documentclass[prb,twocolumn,amsmath,amssymb,floatfix,footinbib,bibnotes,longbibliography]{revtex4-1}
\pdfoutput=1

\usepackage[colorlinks=true,citecolor=blue,linkcolor=magenta]{hyperref}
\usepackage{multirow}
\usepackage{amsmath}
\usepackage{bbm}
\usepackage{graphicx}
\usepackage{dsfont}
\usepackage{changepage}
\usepackage{fancyhdr}
\usepackage{amsthm, amssymb}
\usepackage{array}
\newcolumntype{P}[1]{>{\centering\arraybackslash}p{#1}}
\usepackage[usenames,dvipsnames]{color}
\usepackage[section]{placeins}
\usepackage[T1]{fontenc}
\usepackage[latin9]{inputenc}
\usepackage[active]{srcltx}
\usepackage{color}
\usepackage{amssymb}
\usepackage{esint}
\usepackage[version=4]{mhchem}
\usepackage{comment}
\usepackage{natbib}

\graphicspath{ {Plots} }

\def \mr{\mathrm}

\newcommand{\usenomenclature}{}
\ifdefined\usenomenclature
\usepackage[noprefix]{nomencl}
\fi

\newcommand{\beq}{\begin{equation}}
\newcommand{\eeq}{\end{equation}}

\newcommand{\bcen}{\begin{center}}
\newcommand{\ecen}{\end{center}}
\newcommand{\btab}{\begin{tabular}}
\newcommand{\etab}{\end{tabular}}
\newcommand{\bdes}{\begin{description}}
\newcommand{\edes}{\end{description}}

\newcommand{\bea}{\begin{eqnarray}}
\newcommand{\eea}{\end{eqnarray}}

\newcommand{\bary}{\begin{array}}
\newcommand{\eary}{\end{array}}
\newcommand{\benum}{\begin{enumerate}}
\newcommand{\eenum}{\end{enumerate}}
\newcommand{\bitem}{\begin{itemize}}
\newcommand{\eitem}{\end{itemize}}

\makeatletter

\newcommand{\Rmnum}[1]{\expandafter\@slowromancap\romannumeral #1@}
\makeatother

\newcommand{\ci}{\mathrm{i}}

\newcommand{\tbd}[1]{}

\newcommand{\concept}[1]{}
\newcommand{\figconcept}[1]{}
\newcommand{\eqnconcept}[1]{}
\newcommand{\tabconcept}[1]{}
\newcommand{\appconcept}[1]{}

\newcommand{\G}[2]{G^{#2}_{#1}}

\usepackage{cancel}

\newcommand {\apgt} {\ {\raise-.5ex\hbox{$\buildrel>\over\sim$}}\ }
\newcommand {\aplt} {\ {\raise-.5ex\hbox{$\buildrel<\over\sim$}}\ }

\newcommand {\rem}[1]{}

\def  \w{\omega}
\def  \qinv{Q^{-1}}
\def  \G{\Gamma}
\def  \q0{\frac{\w_k^2}{\G}+z}

\def \titlename {Semiclassical Limit of a Measurement-Induced Transition in Many-Body Chaos in Integrable and Nonintegrable Oscillator Chains}
\def \authornames{Sibaram Ruidas and Sumilan Banerjee}
\def \affiliations{Centre for Condensed Matter Theory, Department of Physics, Indian Institute of Science, Bangalore 560012, India}

\begin{document}
	\title{\titlename}
	\author{\authornames}
	\affiliation{\affiliations}
	\email{sibaramr@iisc.ac.in}
	\email{sumilan@iisc.ac.in}
	\date\today

\begin{abstract}
    We discuss the dynamics of integrable and nonintegrable chains of coupled oscillators under continuous weak position measurements in the semiclassical limit. We show that, in this limit, the dynamics is described by a standard stochastic Langevin equation, and a measurement-induced transition appears as a noise- and dissipation-induced chaotic-to-nonchaotic transition akin to stochastic synchronization. In the nonintegrable chain of anharmonically coupled oscillators, we show that the temporal growth and the ballistic light-cone spread of a classical out-of-time correlator characterized by the Lyapunov exponent and the butterfly velocity, are halted above a noise or below an interaction strength. The Lyapunov exponent and the butterfly velocity both act like  order parameter, vanishing in the nonchaotic phase. In addition, the butterfly velocity exhibits a critical finite-size scaling. For the integrable model, we consider the classical Toda chain and show that the Lyapunov exponent changes nonmonotonically with the noise strength, vanishing at the zero noise limit and above a critical noise, with a maximum at an intermediate noise strength. The butterfly velocity in the Toda chain shows a singular behavior approaching the integrable limit of zero noise strength.
\end{abstract}

\maketitle 
Chaotic-to-nonchaotic transitions play a prominent role in our understanding of the dynamical phase diagram of both quantum and classical systems, appearing in many different contexts such as non-linear dynamics, thermalization and quantum information theory. In quantum many-body systems, a certain kind of chaotic-nonchaotic transitions, dubbed as ``measurement-induced phase transitions" (MIPT) \cite{Skinner2019,Vasseur2020,Choi2020,Yaodong2019,Gullans2020,Nahum2021,Alberton2021,Shengqi2021,Jian2021,Ehud2022,Zabalo2022,Barratt2022} have led to a new paradigm for dynamical phase transitions in recent years. These transitions are characterized by entanglement and chaotic properties of the many-body states and time evolution. On the other hand, a prominent example of transition in chaos in classical dynamical systems \cite{Matsumoto1983,Fahy1992,Maritan1994,Rim2000,Zhou2002,Grassberger1999,Bagnoli1999,Baroni2001,Cencini2001,Pikovsky2002,Ginelli2003,Munoz2003,Bagnoli2006}
are the so-called synchronization transitions (ST)\cite{Pikovsky2001}. In this case, classical trajectories starting from different initial conditions synchronize, i.e. the difference between the trajectories approaches zero with time, when subjected to sufficiently strong common drive, bias, or even random stochastic noise. Can there be some connection between the measurement-induced phase transition in quantum systems and ST in classical systems?

In this Letter, we establish a possible link between MIPT and ST by considering models of interacting particles, whose positions are measured continuously, albeit weakly. We show that, in the semiclassical limit, the dynamics of the system is described by a stochastic Langevin equation where the noise and the dissipation terms are both controlled by the small quantum parameter $\hbar$ and measurement strength. Specifically, we study a nonintegrable oscillator chain, and the classical integrable Toda chain \cite{Toda1967,Henon1974,Flaschka1974-1,Flaschka1974-2}. In both cases, we find a surprising dynamical transition in many-body chaos in the Langevin evolution. The chaotic-to-nonchaotic transition occurs as a function of either interaction or noise (measurement) strength as two classical trajectories starting with slightly different initial conditions synchronize when subjected to identical noise. The transition is similar to the stochastic STs. The latter has been previously studied~\cite{Bagnoli1999,Baroni2001,Cencini2001,Pikovsky2002,Ginelli2003,Munoz2003,Bagnoli2006,Cencini2008,Ginelli2009}, however, only for lattices of coupled non-linear maps, as opposed to interacting Hamiltonian systems employed in our Letter. 

A few recent works have looked into classical analogs of MIPT in cellular automaton \cite{Willsher2022}, kinetically constrain spin systems \cite{Deger2022,Deger2022a}, and semiclassical circuit model \cite{Lyons2022}. In contrast to these works, we derive a direct connection between a quantum measurement dynamics of a Hamiltonian system with the Langevin evolution by analytically taking the semiclassical limit. For interacting quantum systems, MIPTs have been primarily studied in quantum circuits evolving under discrete-time projective and weak measurements \cite{Skinner2019,Vasseur2020,Choi2020,Yaodong2019,Gullans2020,Nahum2021,Alberton2021,Shengqi2021,Jian2021,Ehud2022,Zabalo2022,Barratt2022} as well as continuous weak measurements \cite{Szyniszewski2019,Szyniszewski2020}. The MIPTs in these models are typically characterized by scaling of entanglement entropy with subsystem size, i.e. transition from a volume-law to area-law scaling, in the longtime steady state. 
These MIPTs can often be mapped to phase transitions in some classical statistical mechanics models\cite{Skinner2019,Vasseur2020,Nahum2021,Shengqi2021,Ehud2022,Barratt2022}. However, it is hard to directly take the classical limit of the dynamics in these quantum circuit models. Effects of measurements and MIPTs have also been studied for noninteracting fermions \cite{Cao2019,Lucas2020, Alberton2021}, interacting bosons \cite{Tang2020,Fuji2020,Goto2020}, and Luttinger liquid ground state \cite{Garratt2022}.

We characterize the chaos transition in the Langevin dynamics of the oscillator chains by a classical out-of-time-order correlator (cOTOC) \cite{Subhro2018,Bilitewski2018,Anupam2020,Ruidas2021,Murugan2021,Bilitewski2021,Knolle2021}. The latter is defined by comparing two trajectories that differ by a small amount initially and are subjected to identical noise. In the chaotic phase, we extract a Lyapunov exponent $\lambda_L$ and a butterfly velocity $v_B$, respectively, from the cOTOC. We show the following. (i) $\lambda_L,v_B\to 0$ above a critical noise strength or below an interaction strength for both integrable and nonintegrable chains. (ii) $v_B$ exhibits a critical scaling with system size, whereas $\lambda_L$ shows almost no system-size dependence. The critical exponents extracted from the finite-size scaling of $v_B$ differs from those in the universality classes typically found in stochastic STs in coupled-map lattices (CMLs) \cite{Bagnoli1999,Baroni2001,Ginelli2003,Bagnoli2006,Cencini2008,Ginelli2009}. (iii) For the stochastic dynamics of the integrable Toda chain, $\lambda_L$ changes nonmonotonically with the noise strength, vanishing for zero noise, as well as above a critical noise; $v_B$, on the other hand, shows a singular behavior approaching the integrable limit of zero noise strength.

\begin{figure}[ht]
	\centering
	\includegraphics[width=1.0\linewidth]{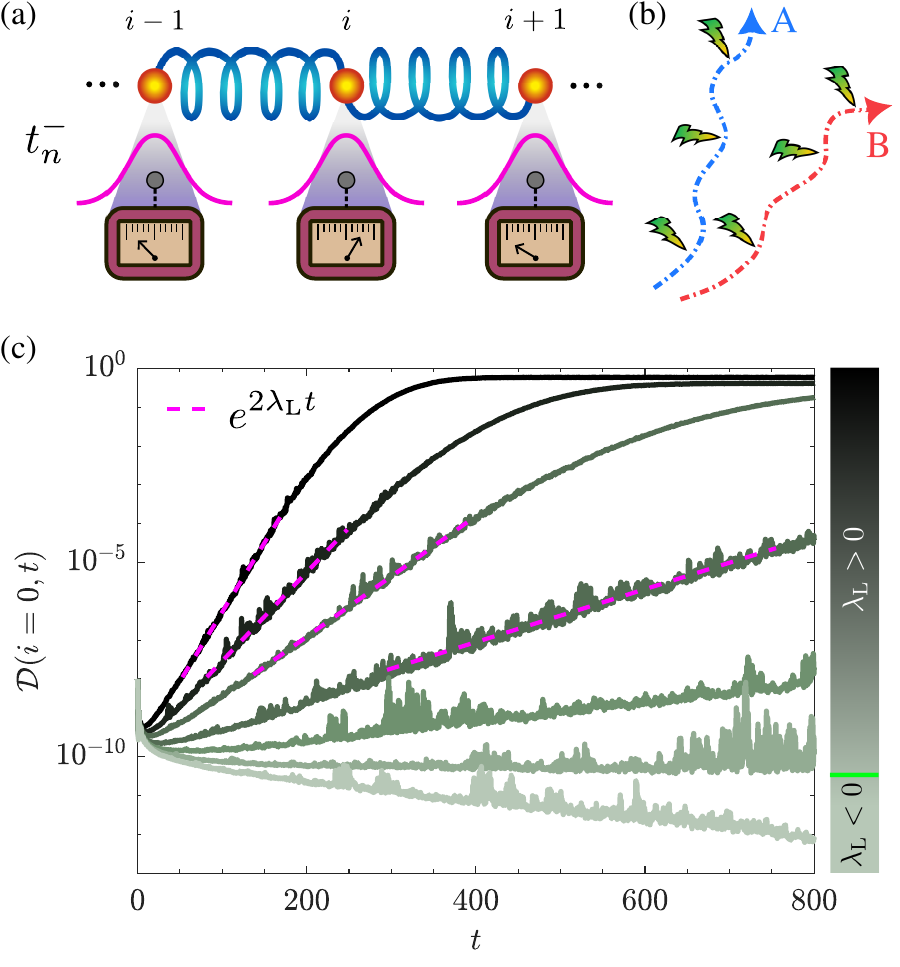}
		\caption{Measurement model and cOTOC. (a) Schematic of the measurement model, where the positions of the coupled oscillators ($i=1,\dots,L$) on a chain are weakly measured at time $t_n$ by meters prepared in Gaussian states just before the measurements. (b) Schematic of two initially nearby classical trajectories, $A$ and $B$, subjected to identical noise realizations. (c) The classical OTOC $\mathcal{D}(i=0,t)$ as a function of $u$ across the chaos transition for $\gamma=0.10$ with $u$ values 0.80 (darkest), 0.60, 0.50, 0.40, 0.35, 0.32 and 0.30 (lightest). As shown by dashed magenta lines, the cOTOC grows exponentially ($\sim e^{2\lambda_Lt}$) for $u>u_c\simeq 0.32$, whereas it decays exponentially for $u<u_c$ in the synchronized phase.} 
	\label{fig:Panel1_Schematics}
\end{figure}

\textit{Quantum measurement model and the semiclassical limit.}---We generalize the well-known model of continuous weak position measurement of a single particle by Caves, and Milburn \cite{Caves1987} to the interacting oscillator chains. The oscillator chain (system) with $i=1,\dots,L$ oscillators and the measurement apparatus (meters) [Fig. \ref{fig:Panel1_Schematics}(a)] are described by the following time-dependent Hamiltonian:
\begin{align}
\mathcal{H}(t)=\mathcal{H}_s+\sum_{i,n} \delta(t-t_n)\hat{x}_i\hat{p}_{in} \label{eq:MeasurementModel}
\end{align}
The Hamiltonian of the system is $\mathcal{H}_s=\sum_i(\hat{p}_i^2/2m)+V(\{\hat{x}_i\})$, where $\hat{x}_i,\hat{p}_i$ are the operators for displacement of the $i$-th oscillator from the equilibrium position and its momentum. We apply periodic boundary conditions. The potential is $V(\{x_i\})=\sum_i v(r_i)$ with $r_i=x_{i+1}-x_i$. We take (i) $v(r)=[(\kappa/2)r^2+(u/4)r^4]$ for the nonintegrable chain with $\kappa$ spring constant and $u$ the strength of the anharmonicity, and (ii) $v(r)=[(a/b)\exp{(-br)}+ar-(a/b)]$ for the integrable Toda chain \cite{Toda1967,Henon1974,Flaschka1974-1,Flaschka1974-2,Abhishek2016} with parameters $a$ and $b$.
The displacement $x_i$ of the $i$-th oscillator is weakly measured by the $in$-th meter at time $t=t_n=n\tau$ at regular intervals of $\tau$. $\hat{p}_{in}$ is the momentum operator of the $in$-th meter, which is in a Gaussian state $\psi(\xi_{in})=(\pi\sigma)^{-1/4}\exp{(-\xi_{in}^2/2\sigma)}$ at $t_n^-$. At $t_n$, the position $\xi_{in}$ of the meter is projectively measured, and its state collapses to a position state $|\xi_{in}\rangle$. The effect of this measurement on the system is described by an operator $\Psi_{i}(\xi_{in})=(\pi\sigma)^{-1/4}\exp{[-(\xi_{in}-\hat{x}_i)^2/2\sigma}]$ acting on the system, as described in detail in the Supplemental Material (SM), Sec.S1. In the continuous measurement limit $\tau\to 0,\sigma\to\infty$ such that $\Delta=\sigma\tau$ is kept fixed \cite{Caves1987}. 

The mean momentum and position of the particle jump by an amount $\propto \xi_{in}$ after each measurement~\cite{Caves1987}, and they can wander far away from the initial values at long times. Thus, to incorporate a feedback mechanism present in any realistic measurement setup \cite{Caves1987} a displacement operator, $D_i(\xi_{in})=\exp{[(\ci/\hbar)\gamma\tau\xi_{in}\hat{p}_i]}$ is applied on the system after the $in$-th measurement, where $\gamma=c_\gamma\sqrt{2\hbar/m\Delta}$, with dimensionless coefficient $c_\gamma$. We do not need to apply a displacement operator for the position due to the periodic boundary condition. 
The feedback mechanism on the momentum leads to dissipation \cite{Caves1987}, as discussed below.

The density matrix of the system at $t_n^+$ is given by $\rho(\{\xi\}_n,t_n^+)=\mathcal{M}(\xi_n)\rho(\{\xi\}_{n-1},t_{n-1}^+)\mathcal{M}^\dagger(\xi_n)$, which depends on the outcomes of all the measurements $\{\xi\}_n$ till $t_n^+$. Here $\mathcal{M}(\xi_n)=\prod_i[D_i(\xi_{in})\Psi_i(\xi_{in})]\exp{(-\ci \mathcal{H}_s\tau/\hbar)}$. For the evolution of an initial pure state, the above time evolution can be written as a quantum state diffusion \cite{Gisin1993,Diosi1998}. Here we write the longtime evolution as a Schwinger-Keldysh (SK) path integral \cite{Kamenev2011} for $\tau\to 0$, i.e. $\mathrm{Tr}[\rho(\{\xi(t)\})]=\int\mathcal{D}x\exp{(\ci S[\{\xi(t)\},x(t)]/\hbar)}$ with the action,
\begin{align}
S&[\{\xi\},x]=\int_{-\infty}^{\infty} dt\sum_{s=\pm}s[\{\sum_i\frac{m}{2}(\dot{x}_i^s)^2+m\gamma\dot{x}_i^s\xi_i\nonumber\\
&+(\ci s\hbar/2\Delta)(x_i^s-\xi_i)^2\}-V(\{x_i^s\})], \label{eq:MeasurementAction}
\end{align}
where $s=\pm$ denotes two branches of the SK contour \cite{Kamenev2011}, $\dot{x}_i=(dx_i^s/dt)$. To take the semiclassical limit of small $\hbar$, we rewrite the above path integral in terms of classical ($x_i^{c}$) and quantum components ($x_i^q$), i.e. $x_i^\pm=x_i^{c}\pm x_i^q$. To capture nontrivial effects of the quantum ($x_i^q$) fluctuations, which act as noise in the semiclassical limit, we need to scale $\Delta\sim \hbar^2$ (SM, Sec.S1). Taking the semiclassical limit in this manner and keeping $\mathcal{O}(1/\sqrt{\hbar})$ and $\mathcal{O}(1)$ terms, we find that a Langevin equation describes the dynamics of the system, 
\begin{align}
\ddot{x}_i^c+\gamma\dot{x}_i^c&=\frac{1}{m}\left[-\frac{\partial V(\{x_i^c\})}{\partial x_i^c}+\eta_i\right],  \label{eq:Dynamics_Gen}
\end{align}
for the classical component $x_i^{c}$, denoted by $x_i$ henceforth. Here the $\eta_i(t)$ is Gaussian random noise that originates from $x_i^q$ and is controlled by the measurement strength $\Delta^{-1}$ such that $\langle \eta_i(t)\eta_j(t')\rangle=2m\gamma T_{\mr{eff}}\delta_{ij}\delta(t-t')$. $T_{\mr{eff}}=(\hbar/4c_\gamma)\sqrt{\hbar/2m\Delta}$, which we denote as $T$ in the rest of the Letter for brevity, is an effective temperature $\sim \sqrt{\hbar}$ that determines the noise strength along with $\gamma$. The latter is the effective dissipation strength $\sim 1/\sqrt{\hbar}$. In the strict classical limit $\hbar\to 0$, $T\to 0$ and $\gamma\to \infty$. As a result, the dissipative term completely dominates, and the system becomes static. The nontrivial semiclassical dynamics results from keeping $\hbar$ small but nonzero. In this limit, the system reaches a longtime steady state described by classical Boltzmann-Gibbs distribution $\sim \exp{[-\mathcal{H}_s(\{x_i,p_i\})/T]}$ determined by the effective temperature. However, the temperature here does not arise from any external baths but solely from the measurement and feedback process.

 \textit{Classical dynamics and cOTOC.}---We study the dynamics [Eq.\eqref{eq:Dynamics_Gen}] of the nonintegrable chain as a function of both $\gamma$ and $u$ for a fixed $T$. The Hamiltonian is trivially integrable and nonchaotic for the harmonic chain ($u = 0$). Any nonzero $u$ makes the model nonintegrable and chaotic. On the contrary, the classical Toda chain is integrable, albeit interacting \cite{Toda1967,Henon1974,Flaschka1974-1,Flaschka1974-2}. We can tune the model from a harmonic limit to a hard sphere limit by changing $a$ and $b$ \cite{Abhishek2016}. We take the parameters in the intermediate regime for the convenience of the numerical simulations. 
 
We numerically simulate Eq.\eqref{eq:Dynamics_Gen} and generate classical trajectories for the nonintegrable and integrable chains using the Gunsteren-Berendsen method \cite{GB1982}; see SM, Sec.S2 for details. We characterize many-body chaos by the following cOTOC, 
 \begin{equation}
     \mathcal{D}(i, t) = \langle [p_{i}^A(t) - p_{i}^B(t)]^{2} \rangle.
     \label{eq:Decorr_p}
 \end{equation} 
Here $A$ and $B$ are two trajectories of the system generated from initial thermal equilibrium configurations $\{x^A_i(0),p_i^A(0)\}$ for $T=1$ with $p_{i}^{B}(0) = p_i^{A}(0) + \delta_{i,0}\varepsilon$ ($\varepsilon = 10^{-4}$); $\langle \cdots \rangle$ denotes average over thermal initial configurations (see SM, Sec.S3 for details). We use $10^5$ initial configurations for all our results. We use identical noise realization for the two copies at each instant of time, i.e., $\eta_i^{{A}}(t) = \eta_i^{{B}}(t)$, as in the earlier studies of stochastic STs and CMLs~\cite{Bagnoli1999,Baroni2001,Cencini2001,Pikovsky2002,Ginelli2003,Munoz2003,Bagnoli2006,Cencini2008,Ginelli2009}. 
  
  \begin{figure}[ht]
	\centering
	\includegraphics[width=1.0\linewidth]{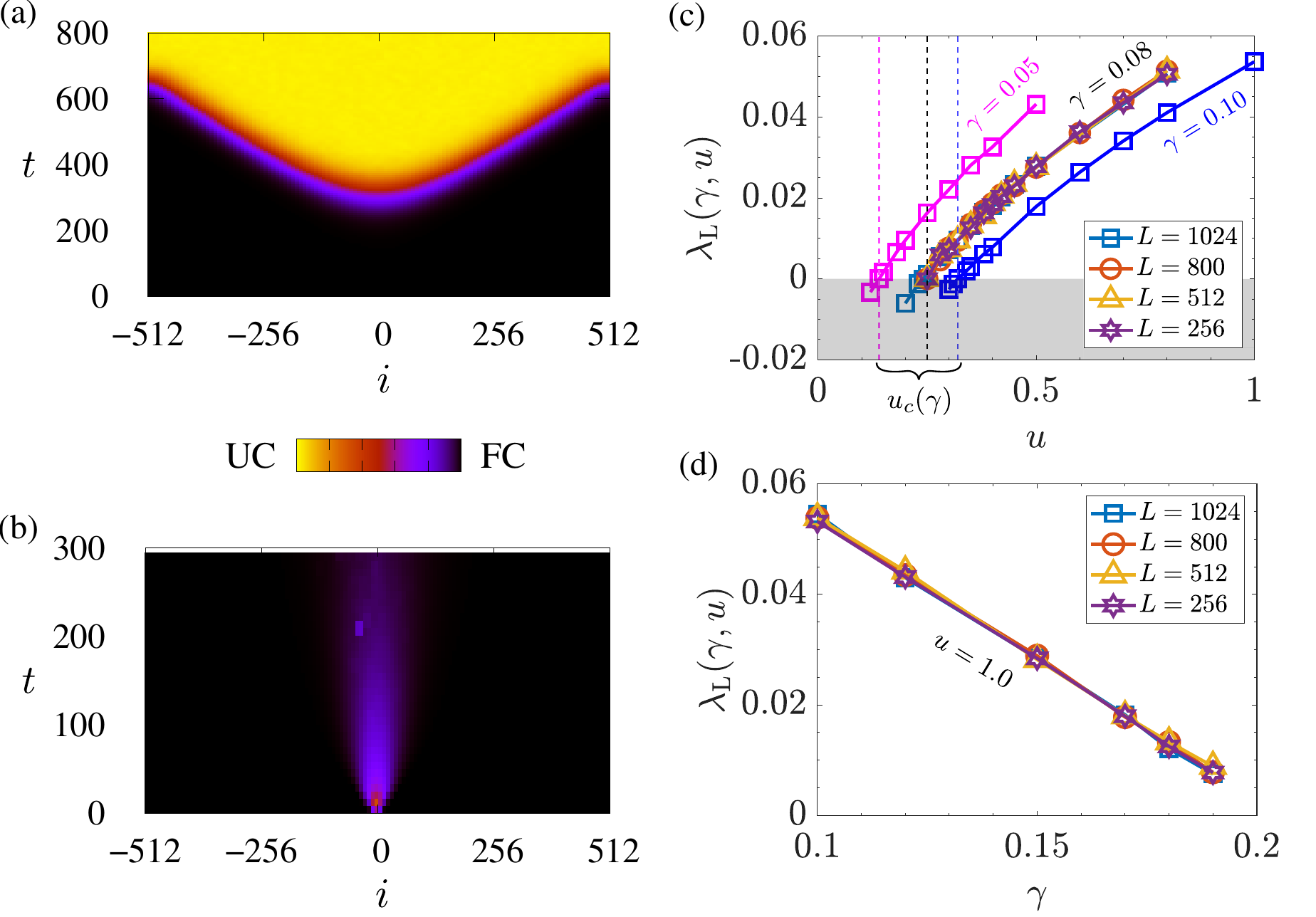}
		\caption{Ballistic light-cone spreading and Lyapunov exponent across the chaos transition. (a) Light cone for $u=0.80>u_c$ and $\gamma=0.10$. The color represents the value of the cOTOC $\mathcal{D}(i,t)$ as function of lattice site $i$ and time $t$, as indicated in the color bar from small $\mathcal{D}(i,t)$ or fully correlated (FC) to large $\mathcal{D}(i,t)$ or uncorrelated (UC). (b) The light-cone spreading ceases in the nonchaotic phase for $u=0.30<u_c$ and $\gamma=0.10$. (c) $\lambda_L$ as a function of $u$ for different $\gamma$s and system sizes $L$. $\lambda_L$ approaches zero at the critical interactions $u_c = 0.14, 0.25$ and $0.32$ (dashed lines) for $\gamma = 0.05, 0.08$ and $0.10$, respectively, for the chaos transition. The shaded region marks $\lambda_L<0$. (d) Similar transition is observed as a function of noise strength $\gamma$ at $\gamma_c \simeq 0.20$ for $u=1$, as shown for different $L$'s.
		} 
	\label{fig:Panel2_transitions}
\end{figure}
 
 \textit{Results.}---For $\gamma=0$, the nonintegrable chain is chaotic, i.e. the cOTOC grows exponentially for any value of $u$, except the harmonic limit $u=0$ (Fig.S2 of SM). However, for $\gamma\neq 0$, the system is chaotic only above a critical value $u_c$ of the interaction, as shown in Fig. \ref{fig:Panel1_Schematics}(c). The cOTOC decays exponentially ($\lambda_L<0$) for $u<u_c$, instead of growing. This is the stochastic ST. Similar transition is seen as a function of noise strength $\gamma$ for a fixed $u\neq 0$ (Fig.S1 of SM). 
The exponential growth is concomitant with a ballistic light cone in cOTOC, whereas the light cone is destroyed in the nonchaotic phase, as shown in Figs. \ref{fig:Panel2_transitions}(a) and \ref{fig:Panel2_transitions}(b). 
 
We extract the Lyapunov exponent $\lambda_L$ from $\mathcal{D}(0,t)\sim \exp{(2\lambda_Lt)}$. The results for $\lambda_L$ as a function of $u$ for a few $\gamma$, and as a function of $\gamma$ for $u=1$, are shown in Figs. \ref{fig:Panel2_transitions}(c) and \ref{fig:Panel2_transitions}(d), for different system sizes $L=256,512,800,1024$. It is evident that $\lambda_L$ approaches zero at a critical value $u_c$ or $\gamma_c$, becoming negative for $u<u_c$ ($\gamma>\gamma_c$), and $\lambda_L$ has little $L$ dependence. Hence the semiclassical limit [Eq.\eqref{eq:Dynamics_Gen}] of the quantum measurement dynamics described by the action in Eq.\eqref{eq:MeasurementAction} indeed yields an ST as a function of the interaction and $\gamma$. Since $\gamma\sim 1/\sqrt{\Delta}$, ST is controlled by the measurement strength $\Delta^{-1}$, which determines how precisely the oscillator positions are measured. As a result, based on its microscopic origin from a quantum measurement model [Eq.\eqref{eq:MeasurementModel}], the ST in this case can be termed as an MIPT, albeit in the semiclassical limit. The transition appears to be continuous, though it is hard to extract $\lambda_L$ accurately close to the transition. 
  
  \begin{figure}[ht]
	\centering
	\includegraphics[width=1.0\linewidth]{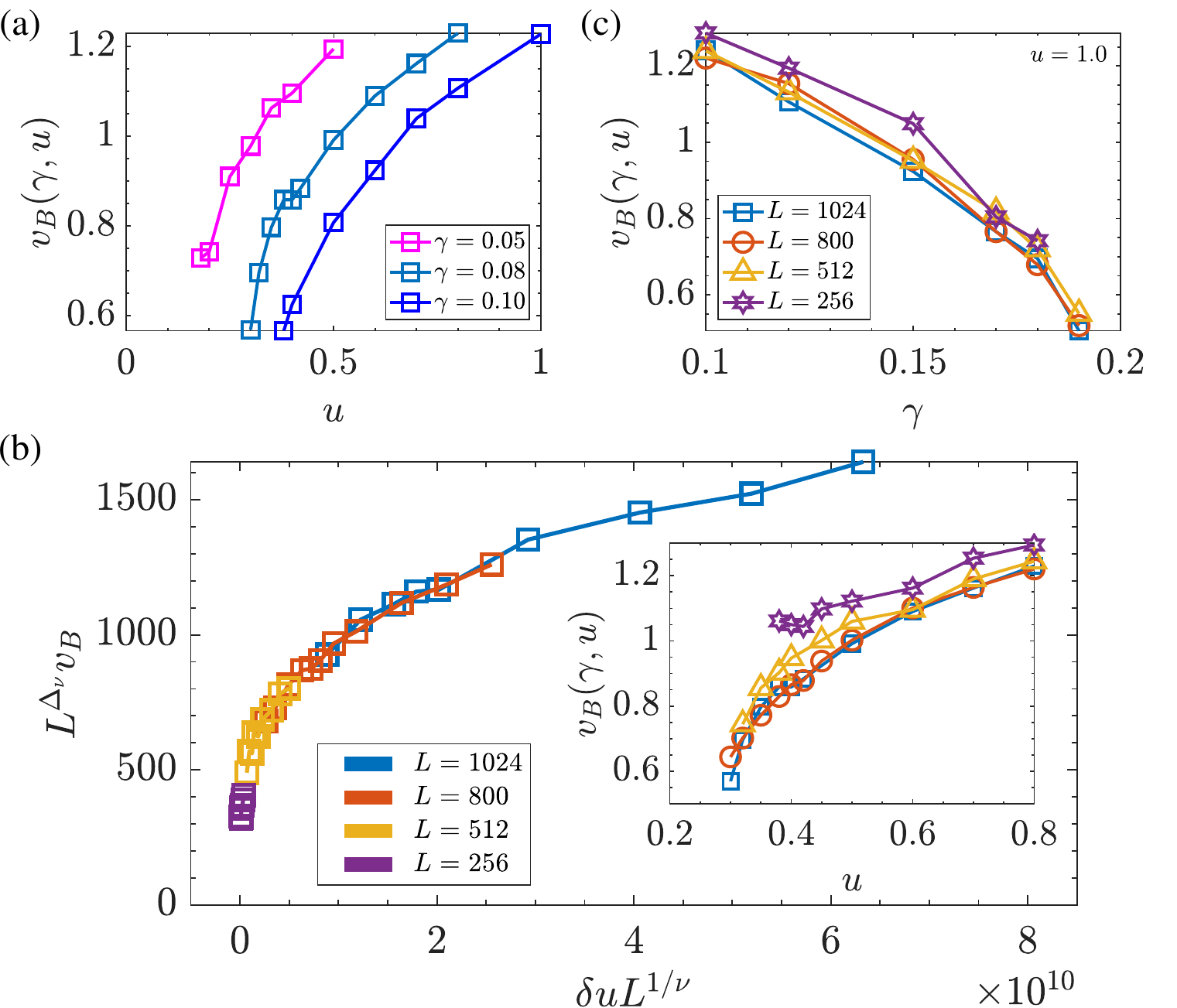}
		\caption{Butterfly velocity and finite-size scaling in the nonintegrable chain. (a) $v_B$ as a function of $u$ for different noise strength $\gamma$. (b) The system size ($L$) dependence of $v_B(u)$ is shown for $\gamma=0.08$ in the inset, and the finite-size scaling collapse is shown in the main panel with exponents $\Delta_v = 1.04$ and $\nu = 0.27$ for $u_c = 0.24$. (c) $v_B$ as a function of $\gamma$ for $u=1.0$ for different $L$s.
		} 
	\label{fig:Panel3_ButtVelo}
\end{figure}

 The ballistic spreading of cOTOC is quantified by extracting butterfly velocity $v_B$, e.g., from the light cones of Fig.\ref{fig:Panel2_transitions}(a) (see SM, Sec.S4 for details). $v_B$ decreases approaching the transition from the chaotic phase, as shown in Figs. \ref{fig:Panel3_ButtVelo}(a) and \ref{fig:Panel3_ButtVelo}(c), as a function of $u$ and $\gamma$, respectively. However, close to the transition, the light cone becomes progressively ill defined and we could not extract $v_B$ all the way up to the transition. Unlike $\lambda_L$, $v_B$ shows perceptible and systematic $L$ dependence [Fig.\ref{fig:Panel3_ButtVelo}(b)(inset)], especially for the transition as function of $u$. Thus we perform a finite-size scaling analysis of the data for $\gamma=0.08$, where we collapse the data for different $L$ and $\delta u=(u-u_c)>0$ using $v_B(u,L)=L^{-\Delta_v}\mathcal{F}((\delta u) L^{1/\nu})$. Here  $\mathcal{F}(x)$ is a scaling function (SM, Sec.S5). Reasonably, good scaling collapse is obtained with $\Delta_v\simeq 1.03\pm 0.03$ and $\nu\simeq 0.30\pm 0.05$, for the range $u_c=0.21$--$0.25$, which is close to the $u_c\simeq 0.25$ obtained from $\lambda_L$ in Fig. \ref{fig:Panel2_transitions}(c). The scaling form implies that for $L\to \infty$, $v_B\sim (\delta u)^{\beta}$ with $\beta=\nu\Delta_v\simeq 0.28$, and a correlation length $\xi$ diverges as $(\delta u)^{-\nu}$ in the chaotic phase. The correlation length exponent $\nu\simeq 0.3$ is different from that for the usual universality classes of STs, such as multiplicative noise or directed percolation, found in earlier studies in CMLs \cite{Bagnoli1999,Baroni2001,Ginelli2003,Bagnoli2006,Cencini2008,Ginelli2009}, cellular automaton \cite{Willsher2022} and kinetically constrained model \cite{Deger2022,Deger2022a}. We note that exponents different from the known universality classes have been found for some cases in previous works on CMLs as well \cite{Baroni2001}.

 The dynamical transition in the stochastic evolution of a nonintegrable oscillator chain is not seen in the usual dynamical properties of a single trajectory. It can only be detected through many-body chaos by comparing two trajectories. To confirm this, 
 we compute the average mean-square displacement (MSD) for the trajectories, i.e. $\langle \Delta q^2(t)\rangle=(1/N)\sum_{i}\langle [x_i(t)-x_i(0)]^2\rangle$ (SM, Sec.S6). For $\gamma=0$, in the harmonic chain ($u=0$) with periodic boundary condition, $\langle \Delta q^2(t)\rangle\sim t$ exhibits a diffusive behavior as shown in ref.\onlinecite{Lee1985}. The diffusive behavior persists for $u\neq 0$ and  $\gamma=0$. However, turning on $\gamma\neq 0$, dynamics becomes subdiffusive with $\langle \Delta q^2(t)\rangle\sim \sqrt{t}$ even for $u=0$. This is well understood in the context of monomer subdiffusion in polymers~\cite{Andrew2010}. Again, for $u\neq 0$ this subdiffusive behavior remains without any change across the ST seen via many-body chaos. We expect the quantum model [Eq.\eqref{eq:MeasurementModel}] to exhibit diffusion in the absence of measurements. It will be interesting to explore the subdiffusive behavior in the presence of measurements in the quantum limit and the connection between diffusion or subdiffusion with entanglement growth \cite{Rakovszky2019,Zhou2020,Znidaric2020}.

\begin{figure}[ht]
	\centering
	\includegraphics[width=1.0\linewidth]{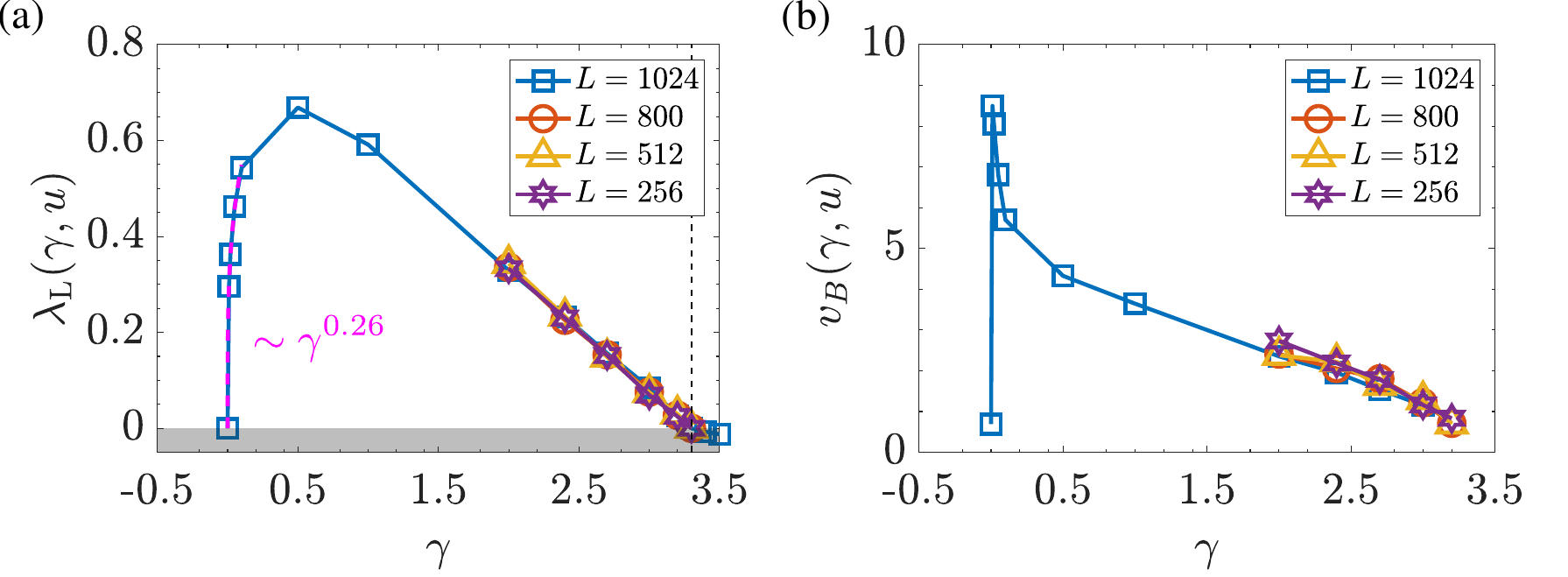}
		\caption{Transitions in many-body chaos in Toda model.  (a) Lyapunov exponent $\lambda_L$ and (b) butterfly velocity $v_B$ as function of noise strength $\gamma$ for different system sizes. For $\gamma\to 0$, $\lambda_L\sim \gamma^{0.26}$ as shown by the dashed line in (a). The shaded region in (a) corresponds to $\lambda_L<0$. The chaos transition occurs around $\gamma_c \simeq 3.30$ (dashed line) for both $\lambda_L$ and $v_B$. 
		} 
	\label{fig:Panle4_TodaResults}
\end{figure}

 We now characterize the many-body chaos in the integrable classical Toda chain. The results for $\lambda_L$ and $v_B$ as a function of $\gamma$ for the Toda chain with $a = 0.07$ and $b = 15.0$ are shown in Figs. \ref{fig:Panle4_TodaResults}(a) and \ref{fig:Panle4_TodaResults}(b) (see SM, Sec. S4 for more details). As expected, the integrable limit with $\gamma = 0$ does not show any exponential growth, implying $\lambda_L=0$. However, the cOTOC still exhibits ballistic spreading in this limit (Fig. S7 of SM), yielding a nonzero $v_B$ as shown in Fig. \ref{fig:Panle4_TodaResults}(b).

 As soon as $\gamma$ becomes nonzero, the dynamics becomes chaotic with both exponential growth and ballistic spreading of cOTOC. As shown in Figs. \ref{fig:Panle4_TodaResults}(a) and \ref{fig:Panle4_TodaResults}(b), the extracted $\lambda_L$ increases\cite{Lam2013,Goldfriend2020} rapidly as $\gamma^{0.3}$ and $v_B$ exhibits a jump near the integrable limit with increasing $\gamma$. Thus, the integrable limit appears singular with respect to $v_B$ for $\gamma\to 0^+$. Further increasing $\gamma$, $v_B$ monotonically decreases and approaches zero at a critical $\gamma=\gamma_c$, indicating a transition to a nonchaotic phase.
In contrast, $\lambda_L$ shows a nonmonotonic dependence on $\gamma$, with a maximum at an intermediate $\gamma$. Nevertheless, $\lambda_L$ eventually vanishes at $\gamma_c$, becoming negative for $\gamma>\gamma_c$, as in the nonintegrable model. Thus, the noise or dissipation, though initially making the integrable model chaotic, eventually destroys chaos due to the stochastic synchronization. The fact that $\lambda_L=0$ for $\gamma=0$ and $\gamma>\gamma_c$, and $\lambda_L>0$ for small $\gamma$ due to the breaking of intigribility\cite{Lam2013,Goldfriend2020}, dictates that $\lambda_L(\gamma)$ is nonmonotonic.

 \textit{Discussions.}--In summary, we show that effective dynamics of the position and momentum of the quantum oscillators under continuous weak position measurements maps to standard stochastic Langevin evolution in the semiclassical limit of small but nonzero $\hbar$. 
 The Langevin dynamics for interacting chains, remarkably, exhibit ST in many-body chaos as a function of noise strength.  The latter is controlled by the measurement strength in the parent quantum measurement model, implying that the ST is an MIPT in the semiclassical limit. 

 The use of stochastic Langevin dynamics might suggest the absence of entanglement in the semicalssical limit. However, this naive inference is not correct. The semicalssical limit of entanglement needs to be taken carefully \cite{Angelo2005,Matzkin2011,Mussardo2022}, where one first obtains an SK path integral for entanglement entropy, e.g., second R\'{e}nyi entropy \cite{Haldar2020}, in the quantum model and then takes the semiclassical limit. This results in effective dynamical equations for entanglement \cite{Banerjee2023} different from Eq.\eqref{eq:Dynamics_Gen}. The latter only describes the effective dynamics of positions and momenta, as typically done in semiclassical approximations \cite{Polkovnikov2010}, and is useful for capturing OTOC \eqref{eq:Decorr_p} in the semiclassical limit. However, the the connection between the growth of OTOC and entanglement has been shown in various situations \cite{Hosur2016,Fan2017,Touil2020}. Hence the ST transition in OTOC suggests an entanglement transition in the semiclassical limit.
 
 The study of the MIPT in the fully quantum limit of our model will be an interesting future direction since systems of interacting oscillators under continuous measurements mimic many realistic open quantum systems. 
 MIPTs have already been shown to exist for continuous weak measurements within more tractable quantum dynamics, like for quantum circuits \cite{Szyniszewski2019,Szyniszewski2020}, and noninteracting fermionic systems \cite{Alberton2021}. Moreover, there are several works \cite{Tang2020,Fuji2020,Goto2020} on the Bose-Hubbard model that show MIPT in the presence of measurements. The oscillator model in our work can be easily mapped to an interacting boson model. Thus, quite generally, MIPT as a function of measurement strength is expected to occur in our models even in the fully quantum limit. 
 
 
 
 We acknowledge useful suggestions and comments by Sriram Ramaswamy and Sitabhra Sinha, and discussions with Subhro Bhattacharjee, Sthitadhi Roy, and Sriram Ganeshan. S.B. acknowledges support from SERB (Grant No CRG/2022/001062), DST, India and QuST, DST, India.

\bibliographystyle{unsrtnat}
\bibliography{supp_chaos} 

\begin{thebibliography}{73}
\providecommand{\natexlab}[1]{#1}
\providecommand{\url}[1]{\texttt{#1}}
\expandafter\ifx\csname urlstyle\endcsname\relax
  \providecommand{\doi}[1]{doi: #1}\else
  \providecommand{\doi}{doi: \begingroup \urlstyle{rm}\Url}\fi

\bibitem[Skinner et~al.(2019)Skinner, Ruhman, and Nahum]{Skinner2019}
Brian Skinner, Jonathan Ruhman, and Adam Nahum.
\newblock Measurement-induced phase transitions in the dynamics of
  entanglement.
\newblock \emph{Phys. Rev. X}, 9:\penalty0 031009, Jul 2019.
\newblock \doi{10.1103/PhysRevX.9.031009}.
\newblock URL \url{https://link.aps.org/doi/10.1103/PhysRevX.9.031009}.

\bibitem[Jian et~al.(2020)Jian, You, Vasseur, and Ludwig]{Vasseur2020}
Chao-Ming Jian, Yi-Zhuang You, Romain Vasseur, and Andreas W.~W. Ludwig.
\newblock Measurement-induced criticality in random quantum circuits.
\newblock \emph{Phys. Rev. B}, 101:\penalty0 104302, Mar 2020.
\newblock \doi{10.1103/PhysRevB.101.104302}.
\newblock URL \url{https://link.aps.org/doi/10.1103/PhysRevB.101.104302}.

\bibitem[Choi et~al.(2020)Choi, Bao, Qi, and Altman]{Choi2020}
Soonwon Choi, Yimu Bao, Xiao-Liang Qi, and Ehud Altman.
\newblock Quantum error correction in scrambling dynamics and
  measurement-induced phase transition.
\newblock \emph{Phys. Rev. Lett.}, 125:\penalty0 030505, Jul 2020.
\newblock \doi{10.1103/PhysRevLett.125.030505}.
\newblock URL \url{https://link.aps.org/doi/10.1103/PhysRevLett.125.030505}.

\bibitem[Li et~al.(2019)Li, Chen, and Fisher]{Yaodong2019}
Yaodong Li, Xiao Chen, and Matthew P.~A. Fisher.
\newblock Measurement-driven entanglement transition in hybrid quantum
  circuits.
\newblock \emph{Phys. Rev. B}, 100:\penalty0 134306, Oct 2019.
\newblock \doi{10.1103/PhysRevB.100.134306}.
\newblock URL \url{https://link.aps.org/doi/10.1103/PhysRevB.100.134306}.

\bibitem[Gullans and Huse(2020)]{Gullans2020}
Michael~J. Gullans and David~A. Huse.
\newblock Dynamical purification phase transition induced by quantum
  measurements.
\newblock \emph{Phys. Rev. X}, 10:\penalty0 041020, Oct 2020.
\newblock \doi{10.1103/PhysRevX.10.041020}.
\newblock URL \url{https://link.aps.org/doi/10.1103/PhysRevX.10.041020}.

\bibitem[Nahum et~al.(2021)Nahum, Roy, Skinner, and Ruhman]{Nahum2021}
Adam Nahum, Sthitadhi Roy, Brian Skinner, and Jonathan Ruhman.
\newblock Measurement and entanglement phase transitions in all-to-all quantum
  circuits, on quantum trees, and in landau-ginsburg theory.
\newblock \emph{PRX Quantum}, 2:\penalty0 010352, Mar 2021.
\newblock \doi{10.1103/PRXQuantum.2.010352}.
\newblock URL \url{https://link.aps.org/doi/10.1103/PRXQuantum.2.010352}.

\bibitem[Alberton et~al.(2021)Alberton, Buchhold, and Diehl]{Alberton2021}
O.~Alberton, M.~Buchhold, and S.~Diehl.
\newblock Entanglement transition in a monitored free-fermion chain: From
  extended criticality to area law.
\newblock \emph{Phys. Rev. Lett.}, 126:\penalty0 170602, Apr 2021.
\newblock \doi{10.1103/PhysRevLett.126.170602}.
\newblock URL \url{https://link.aps.org/doi/10.1103/PhysRevLett.126.170602}.

\bibitem[Sang et~al.(2021)Sang, Li, Zhou, Chen, Hsieh, and Fisher]{Shengqi2021}
Shengqi Sang, Yaodong Li, Tianci Zhou, Xiao Chen, Timothy~H. Hsieh, and
  Matthew~P.A. Fisher.
\newblock Entanglement negativity at measurement-induced criticality.
\newblock \emph{PRX Quantum}, 2:\penalty0 030313, Jul 2021.
\newblock \doi{10.1103/PRXQuantum.2.030313}.
\newblock URL \url{https://link.aps.org/doi/10.1103/PRXQuantum.2.030313}.

\bibitem[Jian et~al.(2021)Jian, Liu, Chen, Swingle, and Zhang]{Jian2021}
Shao-Kai Jian, Chunxiao Liu, Xiao Chen, Brian Swingle, and Pengfei Zhang.
\newblock Measurement-induced phase transition in the monitored
  sachdev-ye-kitaev model.
\newblock \emph{Phys. Rev. Lett.}, 127:\penalty0 140601, Sep 2021.
\newblock \doi{10.1103/PhysRevLett.127.140601}.
\newblock URL \url{https://link.aps.org/doi/10.1103/PhysRevLett.127.140601}.

\bibitem[Block et~al.(2022)Block, Bao, Choi, Altman, and Yao]{Ehud2022}
Maxwell Block, Yimu Bao, Soonwon Choi, Ehud Altman, and Norman~Y. Yao.
\newblock Measurement-induced transition in long-range interacting quantum
  circuits.
\newblock \emph{Phys. Rev. Lett.}, 128:\penalty0 010604, Jan 2022.
\newblock \doi{10.1103/PhysRevLett.128.010604}.
\newblock URL \url{https://link.aps.org/doi/10.1103/PhysRevLett.128.010604}.

\bibitem[Zabalo et~al.(2022)Zabalo, Gullans, Wilson, Vasseur, Ludwig,
  Gopalakrishnan, Huse, and Pixley]{Zabalo2022}
A.~Zabalo, M.~J. Gullans, J.~H. Wilson, R.~Vasseur, A.~W.~W. Ludwig,
  S.~Gopalakrishnan, David~A. Huse, and J.~H. Pixley.
\newblock Operator scaling dimensions and multifractality at
  measurement-induced transitions.
\newblock \emph{Phys. Rev. Lett.}, 128:\penalty0 050602, Feb 2022.
\newblock \doi{10.1103/PhysRevLett.128.050602}.
\newblock URL \url{https://link.aps.org/doi/10.1103/PhysRevLett.128.050602}.

\bibitem[Barratt et~al.(2022)Barratt, Agrawal, Gopalakrishnan, Huse, Vasseur,
  and Potter]{Barratt2022}
Fergus Barratt, Utkarsh Agrawal, Sarang Gopalakrishnan, David~A. Huse, Romain
  Vasseur, and Andrew~C. Potter.
\newblock Field theory of charge sharpening in symmetric monitored quantum
  circuits.
\newblock \emph{Phys. Rev. Lett.}, 129:\penalty0 120604, Sep 2022.
\newblock \doi{10.1103/PhysRevLett.129.120604}.
\newblock URL \url{https://link.aps.org/doi/10.1103/PhysRevLett.129.120604}.

\bibitem[{Matsumoto} and {Tsuda}(1983)]{Matsumoto1983}
K.~{Matsumoto} and I.~{Tsuda}.
\newblock {Noise-induced order}.
\newblock \emph{Journal of Statistical Physics}, 31\penalty0 (1):\penalty0
  87--106, April 1983.
\newblock \doi{10.1007/BF01010923}.

\bibitem[Fahy and Hamann(1992)]{Fahy1992}
S.~Fahy and D.~R. Hamann.
\newblock Transition from chaotic to nonchaotic behavior in randomly driven
  systems.
\newblock \emph{Phys. Rev. Lett.}, 69:\penalty0 761--764, Aug 1992.
\newblock \doi{10.1103/PhysRevLett.69.761}.
\newblock URL \url{https://link.aps.org/doi/10.1103/PhysRevLett.69.761}.

\bibitem[Maritan and Banavar(1994)]{Maritan1994}
Amos Maritan and Jayanth~R. Banavar.
\newblock Chaos, noise, and synchronization.
\newblock \emph{Phys. Rev. Lett.}, 72:\penalty0 1451--1454, Mar 1994.
\newblock \doi{10.1103/PhysRevLett.72.1451}.
\newblock URL \url{https://link.aps.org/doi/10.1103/PhysRevLett.72.1451}.

\bibitem[Rim et~al.(2000)Rim, Hwang, Kim, and Kim]{Rim2000}
Sunghwan Rim, Dong-Uk Hwang, Inbo Kim, and Chil-Min Kim.
\newblock Chaotic transition of random dynamical systems and chaos
  synchronization by common noises.
\newblock \emph{Phys. Rev. Lett.}, 85:\penalty0 2304--2307, Sep 2000.
\newblock \doi{10.1103/PhysRevLett.85.2304}.
\newblock URL \url{https://link.aps.org/doi/10.1103/PhysRevLett.85.2304}.

\bibitem[Zhou and Kurths(2002)]{Zhou2002}
Changsong Zhou and J\"urgen Kurths.
\newblock Noise-induced phase synchronization and synchronization transitions
  in chaotic oscillators.
\newblock \emph{Phys. Rev. Lett.}, 88:\penalty0 230602, May 2002.
\newblock \doi{10.1103/PhysRevLett.88.230602}.
\newblock URL \url{https://link.aps.org/doi/10.1103/PhysRevLett.88.230602}.

\bibitem[Grassberger(1999)]{Grassberger1999}
Peter Grassberger.
\newblock Synchronization of coupled systems with spatiotemporal chaos.
\newblock \emph{Phys. Rev. E}, 59:\penalty0 R2520--R2522, Mar 1999.
\newblock \doi{10.1103/PhysRevE.59.R2520}.
\newblock URL \url{https://link.aps.org/doi/10.1103/PhysRevE.59.R2520}.

\bibitem[Bagnoli et~al.(1999)Bagnoli, Baroni, and Palmerini]{Bagnoli1999}
Franco Bagnoli, Lucia Baroni, and Paolo Palmerini.
\newblock Synchronization and directed percolation in coupled map lattices.
\newblock \emph{Phys. Rev. E}, 59:\penalty0 409--416, Jan 1999.
\newblock \doi{10.1103/PhysRevE.59.409}.
\newblock URL \url{https://link.aps.org/doi/10.1103/PhysRevE.59.409}.

\bibitem[Baroni et~al.(2001)Baroni, Livi, and Torcini]{Baroni2001}
Lucia Baroni, Roberto Livi, and Alessandro Torcini.
\newblock Transition to stochastic synchronization in spatially extended
  systems.
\newblock \emph{Phys. Rev. E}, 63:\penalty0 036226, Feb 2001.
\newblock \doi{10.1103/PhysRevE.63.036226}.
\newblock URL \url{https://link.aps.org/doi/10.1103/PhysRevE.63.036226}.

\bibitem[Cencini and Torcini(2001)]{Cencini2001}
Massimo Cencini and Alessandro Torcini.
\newblock Linear and nonlinear information flow in spatially extended systems.
\newblock \emph{Phys. Rev. E}, 63:\penalty0 056201, Apr 2001.
\newblock \doi{10.1103/PhysRevE.63.056201}.
\newblock URL \url{https://link.aps.org/doi/10.1103/PhysRevE.63.056201}.

\bibitem[Ahlers and Pikovsky(2002)]{Pikovsky2002}
Volker Ahlers and Arkady Pikovsky.
\newblock Critical properties of the synchronization transition in space-time
  chaos.
\newblock \emph{Phys. Rev. Lett.}, 88:\penalty0 254101, Jun 2002.
\newblock \doi{10.1103/PhysRevLett.88.254101}.
\newblock URL \url{https://link.aps.org/doi/10.1103/PhysRevLett.88.254101}.

\bibitem[Ginelli et~al.(2003)Ginelli, Livi, Politi, and Torcini]{Ginelli2003}
F.~Ginelli, R.~Livi, A.~Politi, and A.~Torcini.
\newblock Relationship between directed percolation and the synchronization
  transition in spatially extended systems.
\newblock \emph{Phys. Rev. E}, 67:\penalty0 046217, Apr 2003.
\newblock \doi{10.1103/PhysRevE.67.046217}.
\newblock URL \url{https://link.aps.org/doi/10.1103/PhysRevE.67.046217}.

\bibitem[Mu\~noz and Pastor-Satorras(2003)]{Munoz2003}
Miguel~A. Mu\~noz and Romualdo Pastor-Satorras.
\newblock Stochastic theory of synchronization transitions in extended systems.
\newblock \emph{Phys. Rev. Lett.}, 90:\penalty0 204101, May 2003.
\newblock \doi{10.1103/PhysRevLett.90.204101}.
\newblock URL \url{https://link.aps.org/doi/10.1103/PhysRevLett.90.204101}.

\bibitem[Bagnoli and Rechtman(2006)]{Bagnoli2006}
Franco Bagnoli and Ra\'ul Rechtman.
\newblock Synchronization universality classes and stability of smooth coupled
  map lattices.
\newblock \emph{Phys. Rev. E}, 73:\penalty0 026202, Feb 2006.
\newblock \doi{10.1103/PhysRevE.73.026202}.
\newblock URL \url{https://link.aps.org/doi/10.1103/PhysRevE.73.026202}.

\bibitem[Pikovsky et~al.(2001)Pikovsky, Rosenblum, and Kurths]{Pikovsky2001}
A.~Pikovsky, M.~G. Rosenblum, and J.~Kurths.
\newblock \emph{Synchronization, A Universal Concept in Nonlinear Sciences}.
\newblock Cambridge University Press, Cambridge, 2001.

\bibitem[Toda(1967)]{Toda1967}
Morikazu Toda.
\newblock Wave propagation in anharmonic lattices.
\newblock \emph{Journal of the Physical Society of Japan}, 23\penalty0
  (3):\penalty0 501--506, 1967.

\bibitem[H\'enon(1974)]{Henon1974}
M.~H\'enon.
\newblock Integrals of the toda lattice.
\newblock \emph{Phys. Rev. B}, 9:\penalty0 1921--1923, Feb 1974.
\newblock \doi{10.1103/PhysRevB.9.1921}.
\newblock URL \url{https://link.aps.org/doi/10.1103/PhysRevB.9.1921}.

\bibitem[Flaschka(1974{\natexlab{a}})]{Flaschka1974-1}
Hermann Flaschka.
\newblock The toda lattice. ii. existence of integrals.
\newblock \emph{Physical Review B}, 9\penalty0 (4):\penalty0 1924,
  1974{\natexlab{a}}.

\bibitem[Flaschka(1974{\natexlab{b}})]{Flaschka1974-2}
Hermann Flaschka.
\newblock On the toda lattice. ii: inverse-scattering solution.
\newblock \emph{Progress of Theoretical Physics}, 51\penalty0 (3):\penalty0
  703--716, 1974{\natexlab{b}}.

\bibitem[Cencini et~al.(2008)Cencini, Tessone, and Torcini]{Cencini2008}
M.~Cencini, C.~J. Tessone, and A.~Torcini.
\newblock Chaotic synchronizations of spatially extended systems as
  nonequilibrium phase transitions.
\newblock \emph{Chaos: An Interdisciplinary Journal of Nonlinear Science},
  18\penalty0 (3):\penalty0 037125, 2008.
\newblock \doi{10.1063/1.2945903}.
\newblock URL \url{https://doi.org/10.1063/1.2945903}.

\bibitem[Ginelli et~al.(2009)Ginelli, Cencini, and Torcini]{Ginelli2009}
Francesco Ginelli, Massimo Cencini, and Alessandro Torcini.
\newblock Synchronization of spatio-temporal chaos as an absorbing phase
  transition: a study in 2+1 dimensions.
\newblock \emph{Journal of Statistical Mechanics: Theory and Experiment},
  2009\penalty0 (12):\penalty0 P12018, dec 2009.
\newblock \doi{10.1088/1742-5468/2009/12/p12018}.
\newblock URL \url{https://doi.org/10.1088/1742-5468/2009/12/p12018}.

\bibitem[Willsher et~al.(2022)Willsher, Liu, Moessner, and
  Knolle]{Willsher2022}
Josef Willsher, Shu-Wei Liu, Roderich Moessner, and Johannes Knolle.
\newblock Measurement-induced phase transition in a chaotic classical many-body
  system.
\newblock \emph{Phys. Rev. B}, 106:\penalty0 024305, Jul 2022.
\newblock \doi{10.1103/PhysRevB.106.024305}.
\newblock URL \url{https://link.aps.org/doi/10.1103/PhysRevB.106.024305}.

\bibitem[Deger et~al.(2022{\natexlab{a}})Deger, Roy, and Lazarides]{Deger2022}
Aydin Deger, Sthitadhi Roy, and Achilleas Lazarides.
\newblock Arresting classical many-body chaos by kinetic constraints.
\newblock \emph{Phys. Rev. Lett.}, 129:\penalty0 160601, Oct
  2022{\natexlab{a}}.
\newblock \doi{10.1103/PhysRevLett.129.160601}.
\newblock URL \url{https://link.aps.org/doi/10.1103/PhysRevLett.129.160601}.

\bibitem[Deger et~al.(2022{\natexlab{b}})Deger, Lazarides, and Roy]{Deger2022a}
Aydin Deger, Achilleas Lazarides, and Sthitadhi Roy.
\newblock Constrained dynamics and directed percolation.
\newblock \emph{Phys. Rev. Lett.}, 129:\penalty0 190601, Oct
  2022{\natexlab{b}}.
\newblock \doi{10.1103/PhysRevLett.129.190601}.
\newblock URL \url{https://link.aps.org/doi/10.1103/PhysRevLett.129.190601}.

\bibitem[{Lyons} et~al.(2022){Lyons}, {Choi}, and {Altman}]{Lyons2022}
Anasuya {Lyons}, Soonwon {Choi}, and Ehud {Altman}.
\newblock {A universal crossover in quantum circuits governed by a proximate
  classical error correction transition}.
\newblock \emph{arXiv e-prints}, art. arXiv:2208.02217, August 2022.

\bibitem[Szyniszewski et~al.(2019)Szyniszewski, Romito, and
  Schomerus]{Szyniszewski2019}
M.~Szyniszewski, A.~Romito, and H.~Schomerus.
\newblock Entanglement transition from variable-strength weak measurements.
\newblock \emph{Phys. Rev. B}, 100:\penalty0 064204, Aug 2019.
\newblock \doi{10.1103/PhysRevB.100.064204}.
\newblock URL \url{https://link.aps.org/doi/10.1103/PhysRevB.100.064204}.

\bibitem[Szyniszewski et~al.(2020)Szyniszewski, Romito, and
  Schomerus]{Szyniszewski2020}
M.~Szyniszewski, A.~Romito, and H.~Schomerus.
\newblock Universality of entanglement transitions from stroboscopic to
  continuous measurements.
\newblock \emph{Phys. Rev. Lett.}, 125:\penalty0 210602, Nov 2020.
\newblock \doi{10.1103/PhysRevLett.125.210602}.
\newblock URL \url{https://link.aps.org/doi/10.1103/PhysRevLett.125.210602}.

\bibitem[{Cao} et~al.(2019){Cao}, {Tilloy}, and {De Luca}]{Cao2019}
Xiangyu {Cao}, Antoine {Tilloy}, and Andrea {De Luca}.
\newblock {Entanglement in a fermion chain under continuous monitoring}.
\newblock \emph{SciPost Physics}, 7\penalty0 (2):\penalty0 024, August 2019.
\newblock \doi{10.21468/SciPostPhys.7.2.024}.

\bibitem[Chen et~al.(2020)Chen, Li, Fisher, and Lucas]{Lucas2020}
Xiao Chen, Yaodong Li, Matthew P.~A. Fisher, and Andrew Lucas.
\newblock Emergent conformal symmetry in nonunitary random dynamics of free
  fermions.
\newblock \emph{Phys. Rev. Research}, 2:\penalty0 033017, Jul 2020.
\newblock \doi{10.1103/PhysRevResearch.2.033017}.
\newblock URL \url{https://link.aps.org/doi/10.1103/PhysRevResearch.2.033017}.

\bibitem[Tang and Zhu(2020)]{Tang2020}
Qicheng Tang and W.~Zhu.
\newblock Measurement-induced phase transition: A case study in the
  nonintegrable model by density-matrix renormalization group calculations.
\newblock \emph{Phys. Rev. Res.}, 2:\penalty0 013022, Jan 2020.
\newblock \doi{10.1103/PhysRevResearch.2.013022}.
\newblock URL \url{https://link.aps.org/doi/10.1103/PhysRevResearch.2.013022}.

\bibitem[Fuji and Ashida(2020)]{Fuji2020}
Yohei Fuji and Yuto Ashida.
\newblock Measurement-induced quantum criticality under continuous monitoring.
\newblock \emph{Phys. Rev. B}, 102:\penalty0 054302, Aug 2020.
\newblock \doi{10.1103/PhysRevB.102.054302}.
\newblock URL \url{https://link.aps.org/doi/10.1103/PhysRevB.102.054302}.

\bibitem[Goto and Danshita(2020)]{Goto2020}
Shimpei Goto and Ippei Danshita.
\newblock Measurement-induced transitions of the entanglement scaling law in
  ultracold gases with controllable dissipation.
\newblock \emph{Phys. Rev. A}, 102:\penalty0 033316, Sep 2020.
\newblock \doi{10.1103/PhysRevA.102.033316}.
\newblock URL \url{https://link.aps.org/doi/10.1103/PhysRevA.102.033316}.

\bibitem[{Garratt} et~al.(2022){Garratt}, {Weinstein}, and
  {Altman}]{Garratt2022}
Samuel~J. {Garratt}, Zack {Weinstein}, and Ehud {Altman}.
\newblock {Measurements conspire nonlocally to restructure critical quantum
  states}.
\newblock \emph{arXiv e-prints}, art. arXiv:2207.09476, July 2022.

\bibitem[Das et~al.(2018)Das, Chakrabarty, Dhar, Kundu, Huse, Moessner, Ray,
  and Bhattacharjee]{Subhro2018}
Avijit Das, Saurish Chakrabarty, Abhishek Dhar, Anupam Kundu, David~A. Huse,
  Roderich Moessner, Samriddhi~Sankar Ray, and Subhro Bhattacharjee.
\newblock Light-cone spreading of perturbations and the butterfly effect in a
  classical spin chain.
\newblock \emph{Phys. Rev. Lett.}, 121:\penalty0 024101, Jul 2018.
\newblock \doi{10.1103/PhysRevLett.121.024101}.
\newblock URL \url{https://link.aps.org/doi/10.1103/PhysRevLett.121.024101}.

\bibitem[Bilitewski et~al.(2018)Bilitewski, Bhattacharjee, and
  Moessner]{Bilitewski2018}
Thomas Bilitewski, Subhro Bhattacharjee, and Roderich Moessner.
\newblock Temperature dependence of the butterfly effect in a classical
  many-body system.
\newblock \emph{Phys. Rev. Lett.}, 121:\penalty0 250602, Dec 2018.
\newblock \doi{10.1103/PhysRevLett.121.250602}.
\newblock URL \url{https://link.aps.org/doi/10.1103/PhysRevLett.121.250602}.

\bibitem[Chatterjee et~al.(2020)Chatterjee, Kundu, and Kulkarni]{Anupam2020}
Amit~Kumar Chatterjee, Anupam Kundu, and Manas Kulkarni.
\newblock Spatiotemporal spread of perturbations in a driven dissipative
  duffing chain: An out-of-time-ordered correlator approach.
\newblock \emph{Phys. Rev. E}, 102:\penalty0 052103, Nov 2020.
\newblock \doi{10.1103/PhysRevE.102.052103}.
\newblock URL \url{https://link.aps.org/doi/10.1103/PhysRevE.102.052103}.

\bibitem[Ruidas and Banerjee(2021)]{Ruidas2021}
Sibaram Ruidas and Sumilan Banerjee.
\newblock {Many-body chaos and anomalous diffusion across thermal phase
  transitions in two dimensions}.
\newblock \emph{SciPost Phys.}, 11:\penalty0 087, 2021.
\newblock \doi{10.21468/SciPostPhys.11.5.087}.
\newblock URL \url{https://scipost.org/10.21468/SciPostPhys.11.5.087}.

\bibitem[Murugan et~al.(2021)Murugan, Kumar, Bhattacharjee, and
  Ray]{Murugan2021}
Sugan~Durai Murugan, Dheeraj Kumar, Subhro Bhattacharjee, and Samriddhi~Sankar
  Ray.
\newblock Many-body chaos in thermalized fluids.
\newblock \emph{Phys. Rev. Lett.}, 127:\penalty0 124501, Sep 2021.
\newblock \doi{10.1103/PhysRevLett.127.124501}.
\newblock URL \url{https://link.aps.org/doi/10.1103/PhysRevLett.127.124501}.

\bibitem[Bilitewski et~al.(2021)Bilitewski, Bhattacharjee, and
  Moessner]{Bilitewski2021}
Thomas Bilitewski, Subhro Bhattacharjee, and Roderich Moessner.
\newblock Classical many-body chaos with and without quasiparticles.
\newblock \emph{Phys. Rev. B}, 103:\penalty0 174302, May 2021.
\newblock \doi{10.1103/PhysRevB.103.174302}.
\newblock URL \url{https://link.aps.org/doi/10.1103/PhysRevB.103.174302}.

\bibitem[Liu et~al.(2021)Liu, Willsher, Bilitewski, Li, Smith, Christensen,
  Moessner, and Knolle]{Knolle2021}
Shu-Wei Liu, J.~Willsher, T.~Bilitewski, Jin-Jie Li, A.~Smith, K.~Christensen,
  R.~Moessner, and J.~Knolle.
\newblock Butterfly effect and spatial structure of information spreading in a
  chaotic cellular automaton.
\newblock \emph{Phys. Rev. B}, 103:\penalty0 094109, Mar 2021.
\newblock \doi{10.1103/PhysRevB.103.094109}.
\newblock URL \url{https://link.aps.org/doi/10.1103/PhysRevB.103.094109}.

\bibitem[Caves and Milburn(1987)]{Caves1987}
Carlton~M. Caves and G.~J. Milburn.
\newblock Quantum-mechanical model for continuous position measurements.
\newblock \emph{Phys. Rev. A}, 36:\penalty0 5543--5555, Dec 1987.
\newblock \doi{10.1103/PhysRevA.36.5543}.
\newblock URL \url{https://link.aps.org/doi/10.1103/PhysRevA.36.5543}.

\bibitem[Kundu and Dhar(2016)]{Abhishek2016}
Aritra Kundu and Abhishek Dhar.
\newblock Equilibrium dynamical correlations in the toda chain and other
  integrable models.
\newblock \emph{Phys. Rev. E}, 94:\penalty0 062130, Dec 2016.
\newblock \doi{10.1103/PhysRevE.94.062130}.
\newblock URL \url{https://link.aps.org/doi/10.1103/PhysRevE.94.062130}.

\bibitem[Gisin and Percival(1993)]{Gisin1993}
Nicolas Gisin and Ian~C Percival.
\newblock Quantum state diffusion, localization and quantum dispersion entropy.
\newblock \emph{Journal of Physics A: Mathematical and General}, 26\penalty0
  (9):\penalty0 2233, 1993.

\bibitem[Di\'osi et~al.(1998)Di\'osi, Gisin, and Strunz]{Diosi1998}
L.~Di\'osi, N.~Gisin, and W.~T. Strunz.
\newblock Non-markovian quantum state diffusion.
\newblock \emph{Phys. Rev. A}, 58:\penalty0 1699--1712, Sep 1998.
\newblock \doi{10.1103/PhysRevA.58.1699}.
\newblock URL \url{https://link.aps.org/doi/10.1103/PhysRevA.58.1699}.

\bibitem[Kamenev(2011)]{Kamenev2011}
Alex Kamenev.
\newblock \emph{Field theory of non-equilibrium systems}.
\newblock Cambridge University Press, 2011.

\bibitem[Van~Gunsteren and Berendsen(1982)]{GB1982}
WF~Van~Gunsteren and HJC Berendsen.
\newblock Algorithms for brownian dynamics.
\newblock \emph{Molecular Physics}, 45\penalty0 (3):\penalty0 637--647, 1982.

\bibitem[Florencio and Lee(1985)]{Lee1985}
J.~Florencio and M.~Howard Lee.
\newblock Exact time evolution of a classical harmonic-oscillator chain.
\newblock \emph{Phys. Rev. A}, 31:\penalty0 3231--3236, May 1985.
\newblock \doi{10.1103/PhysRevA.31.3231}.
\newblock URL \url{https://link.aps.org/doi/10.1103/PhysRevA.31.3231}.

\bibitem[Weber et~al.(2010)Weber, Theriot, and Spakowitz]{Andrew2010}
Stephanie~C. Weber, Julie~A. Theriot, and Andrew~J. Spakowitz.
\newblock Subdiffusive motion of a polymer composed of subdiffusive monomers.
\newblock \emph{Phys. Rev. E}, 82:\penalty0 011913, Jul 2010.
\newblock \doi{10.1103/PhysRevE.82.011913}.
\newblock URL \url{https://link.aps.org/doi/10.1103/PhysRevE.82.011913}.

\bibitem[Rakovszky et~al.(2019)Rakovszky, Pollmann, and von
  Keyserlingk]{Rakovszky2019}
Tibor Rakovszky, Frank Pollmann, and C.~W. von Keyserlingk.
\newblock Sub-ballistic growth of r\'enyi entropies due to diffusion.
\newblock \emph{Phys. Rev. Lett.}, 122:\penalty0 250602, Jun 2019.
\newblock \doi{10.1103/PhysRevLett.122.250602}.
\newblock URL \url{https://link.aps.org/doi/10.1103/PhysRevLett.122.250602}.

\bibitem[Zhou and Ludwig(2020)]{Zhou2020}
Tianci Zhou and Andreas W.~W. Ludwig.
\newblock Diffusive scaling of r\'enyi entanglement entropy.
\newblock \emph{Phys. Rev. Res.}, 2:\penalty0 033020, Jul 2020.
\newblock \doi{10.1103/PhysRevResearch.2.033020}.
\newblock URL \url{https://link.aps.org/doi/10.1103/PhysRevResearch.2.033020}.

\bibitem[{{\v{Z}}nidari{\v{c}}}(2020)]{Znidaric2020}
Marko {{\v{Z}}nidari{\v{c}}}.
\newblock {Entanglement growth in diffusive systems}.
\newblock \emph{Communications Physics}, 3\penalty0 (1):\penalty0 100, June
  2020.
\newblock \doi{10.1038/s42005-020-0366-7}.

\bibitem[{Nguyen Thu Lam} and {Kurchan}(2013)]{Lam2013}
Khanh-Dang {Nguyen Thu Lam} and Jorge {Kurchan}.
\newblock {Stochastic perturbation of integrable systems: a window to weakly
  chaotic systems}.
\newblock \emph{arXiv e-prints}, art. arXiv:1305.4503, May 2013.

\bibitem[Goldfriend and Kurchan(2020)]{Goldfriend2020}
Tomer Goldfriend and Jorge Kurchan.
\newblock Quasi-integrable systems are slow to thermalize but may be good
  scramblers.
\newblock \emph{Phys. Rev. E}, 102:\penalty0 022201, Aug 2020.
\newblock \doi{10.1103/PhysRevE.102.022201}.
\newblock URL \url{https://link.aps.org/doi/10.1103/PhysRevE.102.022201}.

\bibitem[Angelo and Furuya(2005)]{Angelo2005}
Renato~M. Angelo and K.~Furuya.
\newblock Semiclassical limit of the entanglement in closed pure systems.
\newblock \emph{Phys. Rev. A}, 71:\penalty0 042321, Apr 2005.
\newblock \doi{10.1103/PhysRevA.71.042321}.
\newblock URL \url{https://link.aps.org/doi/10.1103/PhysRevA.71.042321}.

\bibitem[Matzkin(2011)]{Matzkin2011}
A.~Matzkin.
\newblock Entanglement in the classical limit: Quantum correlations from
  classical probabilities.
\newblock \emph{Phys. Rev. A}, 84:\penalty0 022111, Aug 2011.
\newblock \doi{10.1103/PhysRevA.84.022111}.
\newblock URL \url{https://link.aps.org/doi/10.1103/PhysRevA.84.022111}.

\bibitem[Mussardo and Viti(2022)]{Mussardo2022}
G.~Mussardo and J.~Viti.
\newblock $\ensuremath{\hbar}\ensuremath{\rightarrow}0$ limit of the
  entanglement entropy.
\newblock \emph{Phys. Rev. A}, 105:\penalty0 032404, Mar 2022.
\newblock \doi{10.1103/PhysRevA.105.032404}.
\newblock URL \url{https://link.aps.org/doi/10.1103/PhysRevA.105.032404}.

\bibitem[Haldar et~al.(2020)Haldar, Bera, and Banerjee]{Haldar2020}
Arijit Haldar, Surajit Bera, and Sumilan Banerjee.
\newblock R\'enyi entanglement entropy of fermi and non-fermi liquids:
  Sachdev-ye-kitaev model and dynamical mean field theories.
\newblock \emph{Phys. Rev. Res.}, 2:\penalty0 033505, Sep 2020.
\newblock \doi{10.1103/PhysRevResearch.2.033505}.
\newblock URL \url{https://link.aps.org/doi/10.1103/PhysRevResearch.2.033505}.

\bibitem[Banerjee(2023)]{Banerjee2023}
S.~Banerjee, 2023.
\newblock unpublished.

\bibitem[{Polkovnikov}(2010)]{Polkovnikov2010}
Anatoli {Polkovnikov}.
\newblock {Phase space representation of quantum dynamics}.
\newblock \emph{Annals of Physics}, 325\penalty0 (8):\penalty0 1790--1852,
  August 2010.
\newblock \doi{10.1016/j.aop.2010.02.006}.

\bibitem[{Hosur} et~al.(2016){Hosur}, {Qi}, {Roberts}, and
  {Yoshida}]{Hosur2016}
Pavan {Hosur}, Xiao-Liang {Qi}, Daniel~A. {Roberts}, and Beni {Yoshida}.
\newblock {Chaos in quantum channels}.
\newblock \emph{Journal of High Energy Physics}, 2016:\penalty0 4, February
  2016.
\newblock \doi{10.1007/JHEP02(2016)004}.

\bibitem[{Fan} et~al.(2017){Fan}, {Zhang}, {Shen}, and {Zhai}]{Fan2017}
Ruihua {Fan}, Pengfei {Zhang}, Huitao {Shen}, and Hui {Zhai}.
\newblock {Out-of-time-order correlation for many-body localization}.
\newblock \emph{Science Bulletin}, 62\penalty0 (10):\penalty0 707--711, May
  2017.
\newblock \doi{10.1016/j.scib.2017.04.011}.

\bibitem[{Touil} and {Deffner}(2020)]{Touil2020}
Akram {Touil} and Sebastian {Deffner}.
\newblock {Quantum scrambling and the growth of mutual information}.
\newblock \emph{Quantum Science and Technology}, 5\penalty0 (3):\penalty0
  035005, July 2020.
\newblock \doi{10.1088/2058-9565/ab8ebb}.

\end{thebibliography}
 
\clearpage
\newpage
\def\makeSM{1}
\ifdefined\makeSM

\appendix
\renewcommand{\appendixname}{}
\renewcommand{\thesection}{{S\arabic{section}}}
\renewcommand{\theequation}{\thesection.\arabic{equation}}
\renewcommand{\thefigure}{S\arabic{figure}}
\setcounter{page}{1}
\setcounter{figure}{0}
\setcounter{equation}{0}

\widetext

\centerline{\bf Supplemental Material}
\centerline{\bf for}
\begin{center}
\bf Semiclassical limit of a measurement-induced transition in many-body chaos \\ in integrable and nonintegrable oscillator chains
\end{center}
\centerline{by \authornames}
\affiliation{\affiliations}
\fi 

\def  \qinv{Q^{-1}}
\def  \q0{\frac{\w_k^2}{\G}+z}

\newcommand{\lin}{linspace}
\newcommand{\m}{\Delta_0}
\newcommand{\M}{\Delta}
\newcommand{\g}{Q}
\newcommand{\ginv}{\big(Q^{-1}\big)_{ab}}
\newcommand{\qr}{q_{\text{reg}}} 
\newcommand{\sr}{\Sigma_{\text{reg}}}
\newcommand{\qt}{\tilde{q}_{{EA}}}
\newcommand{\bfig}{\begin{figure}[H]\centering}
\newcommand{\efig}{\end{figure}}
\newcommand{\s}{\hspace{0.5cm }} 
\newcommand{\sh}{\hspace{0.25cm }} 

\section{Measurement Model}\label{sec:Model_S}
For the model of Eq.\eqref{eq:MeasurementModel} and the measurement protocol discussed in the main text, before the $n$-th measurements, the $n$-th meters and the system are described by the density matrix
\begin{align}
\rho_{tot}(\{\xi\}_{n-1},t_{n}^{-})= & \otimes_{i}|\psi_{in}\rangle\langle\psi_{in}|\otimes e^{-\ci\mathcal{H}_s\tau/\hbar}\rho(\{\xi\}_{n-1},t_{n-1}^{+})e^{\ci\mathcal{H}_s\tau/\hbar},
\end{align}
where $\rho(\{\xi\}_{n-1},t_{n-1}^{+})$ is density matrix of the system after $(n-1)$-th measurements and it depends on the outcomes $\{\xi\}_{n-1}$ of all the measurements till $t_{n-1}^+$. In the following we simply write the density matrix as $\rho(t)$ for brevity. After the $n$-th measurement the $n$-th meters are projected into one of the position states $\xi_{in}$. As a result, the state of the combined system just after the $n$-th measurements is
\begin{align}
\rho_{tot}(t_{n}^{+})= & \mathcal{P}_{\xi_{n}}\left[e^{-(\ci/\hbar)\sum_{i}\hat{x}_{i}\hat{p}_{in}}\otimes_{i}|\psi_{in}\rangle\langle\psi_{in}|\otimes e^{-\ci\mathcal{H}_s\tau/\hbar}\rho(t_{n-1}^{+})e^{\ci\mathcal{H}_s\tau/\hbar}e^{(\ci/\hbar)\sum_{i}\hat{x}_{i}\hat{p}_{in}}\right]\mathcal{P}_{\xi_{n}}
\end{align}
where $\mathcal{P}_{\xi_{n}}=\otimes_{i}|\xi_{in}\rangle\langle\xi_{in}|$ is the projection operator applied on the $in$-th meter. As a result the state of the system after the measurement is 
\begin{align}
\rho(t_{n}^{+})= & \mathrm{tr}\left[\rho_{tot}(t_{n}^{+})\right]
\end{align}
where the trace is only over the meters. Thus we get
\begin{align}
\rho(t_{n}^{+})= & \prod_{i}{\Psi}_{i}(\xi_{in})e^{-\ci\mathcal{H}_s\tau/\hbar}\rho(t_{n-1}^{+})e^{\ci\mathcal{H}_s\tau/\hbar}\prod_{i}{\Psi}_{i}^{\dagger}(\xi_{in})
\end{align}
where
\begin{align}
\Psi_{i}(\xi_{in})=\langle\xi_{in}|\psi_{in}\rangle= & (\pi\sigma)^{-1/4}e^{-(\xi_{in}-\hat{x}_{i})^{2}/2\sigma}
\end{align}
 Following Ref.\onlinecite{Caves1987}, and as discussed in the main text, we incorporate a feedback controlling the drift of the momentum of the oscillators by applying displacement operators
\begin{align}
D_{i}(\xi_{in})= & e^{-(\ci/\hbar)\gamma\tau\xi_{in}\hat{p}_{i}},
\end{align}
 after each measurement, with $\gamma=c_\gamma\sqrt{2\hbar/(m\Delta)}$, and $\Delta=\sigma \tau$. Thus the state of the system becomes
 \begin{align}
\rho_n&\equiv\rho(t_{n}^{+})= \mathcal{M}(\xi_n)\rho_{n-1}\mathcal{M}^\dagger(\xi_n) \label{eq:DensityMatrixEvolution_S}\\
\mathcal{M}(\xi_n)&\equiv\prod_{i}\left[D_{i}(\xi_{in})\Psi_{i}(\xi_{in})\right]e^{-\ci\mathcal{H}_s\tau/\hbar}
\end{align}

We now write a path integral representation of the time evolution on the Schwinger-Keldysh (SK) contour \cite{Kamenev2011} by using
\begin{align}
\rho_{n}= & \int dx_{n}^{+}dx_{n}^{-}dx_{n-1}^{+}dx_{n-1}^{-}|x_{n}^{+}\rangle\langle x_{n}^{-}|\langle x_{n}^{+}|\mathcal{M}(\xi_{n})|x_{n-1}^{+}\rangle\langle x_{n-1}^{-}|\mathcal{M}^{\dagger}(\xi_{n})|x_{n}^{-}\rangle\langle x_{n-1}^{+}|\rho_{n-1}|x_{n-1}^{-}\rangle,
\end{align}
where we have omitted the lattice index $i$ for brevity. The matrix element is
\begin{align}
\langle x_{n}^{+}|\mathcal{M}(\xi_{n})|x_{n-1}^{+}\rangle\propto & e^{-(\xi_{n}-x_{n}^{+})^{2}/2\sigma} \exp\left[\frac{\ci}{\hbar}\left\{ \frac{m}{2}(\partial_{t}x_{n-1}^{+})^{2}+m\gamma\xi_{n}\partial_{t}x_{n-1}^{+}-V(x_{n-1}^{+})\right\} \right],
\end{align}
where $\partial_t x_{n-1}^+=(x_n^+-x_{n-1}^+)/\tau$. Finally, taking the continuum limit, we write the SK generating function for the dynamics at time $t_f$ for an initial density matrix $\rho_{0}$ at time $t_0$ as
\begin{align}
\mathrm{Tr}[\rho(\{\xi\},t_f)]= & \int\mathcal{D}x e^{(\ci/\hbar)S[\{\xi\},x]}\langle x(t_0+)|\rho_{0}|x(t_0-)\rangle,
\end{align}
with the action
\begin{align}
S[\{\xi\},x]= & \int_{t_0}^{t_f}dt\sum_{i,s}s\left[\frac{m}{2}(\dot{x}_{i}^{s})^{2}+m\gamma\xi_{i}(t)\dot{x}_{i}^{s}+\frac{\ci s\hbar}{2\Delta}(x_{i}^{s}(t)-\xi_{i}(t))^{2}\right]-\int_{-\infty}^{\infty}dt\sum_{s}sV(\{x_{i}^{s}(t)\}) \label{eq:KeldyshAction_S},
\end{align}
where $s=\pm$ refers to forward and backward branches of the SK contour. Since we are interested in the non-equilibrium steady state (NESS), we eventually take $t_0\to -\infty$ so that the dependence on the initial density matrix drops out from the path integral.

To take the semiclassical limit, we define classical and quantum components as
\begin{align}
x_{i}^{\pm}= & x_{i}^{c}\pm x_{i}^{q},
\end{align}
and write the path integral as 
\begin{align}
&\rho(\{\xi\},t)= \int\mathcal{D}x^{c}\mathcal{D}x^{q}e^{\ci S[\{\xi\},x]}\nonumber\\
&S=\int_{-\infty}^{t_f}dt\sum_{i}2\left[-mx_{i}^{q}\ddot{x}_{i}^{c}+m\gamma\hbar\dot{x}_{i}^{q}\widetilde{\xi}_{i}-m\gamma x_{i}^{q}\dot{x}_{i}^{c}-x_{i}^{q}\frac{\partial V}{\partial x_{i}^{c}}+\hbar^2\mathcal{O}((x_{i}^{q})^{3})\right]+\frac{\ci\hbar^2}{\Delta}\int_{-\infty}^{t_f}dt\sum_{i}\left[\widetilde{\xi}_{i}^{2}+(x_{i}^{q})^{2}\right] \label{eq:MeasurementAction_S}
\end{align}
In the above, we have expanded in $x^{q}$ around $x^{c}$, defined $\widetilde{\xi}_i=\xi_i-x_i^c$, and scaled $x^{q}$ and $\widetilde{\xi}$ as $(x_{i}^{q}/\hbar,\widetilde{\xi}_{i}/\hbar)\to (x_{i}^{q},\widetilde{\xi}_{i})$. We have also used an integration by parts, $\int dt \dot{x}_i^qx_i^c=x_i^qx_i^c|_{-\infty}^{t_f}-\int dt x_i^q \dot{x}_i^c$ and neglected the boundary terms above.

Now to take the semiclassical limit of small $\hbar$, we take $\hbar^{2}/\Delta$ a constant in the limit $\hbar\to 0$, i.e. $\Delta\sim\hbar^{2}$, so that 
\begin{align}
x^{c},\tilde{\xi},x^{q}\sim & \mathcal{O}(1).
\end{align}
Thus we ensure that $\widetilde{\xi}$ and $x^{q}$ Gaussian distributed with width $\sim\mathcal{O}(1)$ in the semiclassical limit. Since $\gamma\sim\sqrt{\hbar/\Delta}\sim1/\sqrt{\hbar}$, we get for the different terms in the action of Eq.\eqref{eq:MeasurementAction_S},
$x^{q}\partial_{t}^{2}x^{c}\sim \mathcal{O}(1)$, $\gamma\hbar\partial_{t}x^{q}\widetilde{\xi} \sim\mathcal{O}(\hbar^{1/2})$, $
\gamma x^{q}\partial_{t}x^{c}\sim \mathcal{O}(1/\sqrt{\hbar})$, and 
$x^{q}(\partial V/\partial x^{c})\sim \mathcal{O}(1)$.

Keeping the leading and the next to leading order terms, $\mathcal{O}(1/\sqrt{\hbar}$)
and $\mathcal{O}(1)$, we get
\begin{align}
S\simeq & \int_{-\infty}^{t_f}dt\sum_{i}2\left[-mx_{i}^{q}\ddot{x}_{i}^{c}-m\gamma x_{i}^{q}\dot{x}_{i}^{c}-x_{i}^{q}\frac{\partial V}{\partial x_{i}^{c}}\right]+\frac{\ci\hbar^{2}}{\Delta}\int_{-\infty}^{t_f}dt\sum_{i}[\widetilde{\xi}_i^2+(x_{i}^{q})^{2}]
\end{align}

We now use the Hubbard Stratonovich identity, 
\begin{align}
e^{-\frac{\hbar^{2}}{\Delta}\int_{-\infty}^{t_f}dt\sum_{i}(x_{i}^{q})^{2}}\propto & \int\mathcal{D}\eta e^{-\frac{\Delta}{\hbar^{2}}\int_{-\infty}^{t_f}dt\sum_{i}\eta_{i}^{2}+\ci\int_{-\infty}^{t_f}dt\sum_{i}2\eta_{i}x_{i}^{q}},
\end{align}
to write
\begin{align}
\rho(t)\propto & \int\mathcal{D}x^{c}\mathcal{D}\eta e^{-\frac{\hbar^2}{\Delta}\int_{-\infty}^{t_f}dt\sum_{i}\widetilde{\xi}_{i}^{2}} e^{-\frac{\Delta}{\hbar^{2}}\int_{-\infty}^{t_f}dt\sum_{i} \eta_{i}^{2}}\int\mathcal{D}x^{q}\exp\left[-2\ci\int_{-\infty}^{t_f}dt\sum_{i}x_{i}^{q}\left(m\ddot{x}_{i}^{c}+m\gamma\dot{x}_{i}^{c}+\frac{\partial V}{\partial x_{i}^{c}}-\eta_{i}\right)\right]\\
\propto & \int\mathcal{D}\eta e^{-\frac{\hbar^2}{\Delta}\int_{-\infty}^{t_f}dt\sum_{i}\widetilde{\xi}_{i}^{2}}e^{-\frac{\Delta}{\hbar^{2}}\int_{-\infty}^{t_f}dt\sum_{i} \eta_{i}^{2}}\int\mathcal{D}x^{c}\prod_{it}\delta\left(m\ddot{x}_{i}^{c}+m\gamma\dot{x}_{i}^{c}+\frac{\partial V}{\partial x_{i}^{c}}-\eta_{i}\right)
\end{align}
The above leads to the classical stochastic Langevin equation for the effective dynamics of the classical component $x_i\equiv x_i^c$
\begin{align}
m\frac{d^{2}x_{i}}{dt^{2}}= & -m\gamma\frac{dx_{i}}{dt}-\frac{\partial V}{\partial x_{i}}+\eta_{i} \label{eq:LangevinEqn_S}
\end{align}
with a noise $\eta_{i}$, which originates from quantum ($x_i^q$) fluctuations. The noise has a Gaussian distribution controlled by the measurement strength $\Delta^{-1}$, such that
\begin{align}
\langle\eta_{i}(t)\eta_{j}(t')\rangle= &\frac{\hbar^{2}}{2\Delta}\delta_{ij}\delta(t-t')\hspace{1em}
\end{align}
Hence, we can define an effective temperature $T_{eff}$ by demanding a fluctuation
dissipation relation between the noise and damping terms, i.e.
\begin{align}
T_{eff}= & \frac{\hbar^{2}}{4m\gamma\Delta}=\frac{\hbar}{4c_\gamma}\sqrt{\frac{\hbar}{2m\Delta}}\sim\sqrt{\hbar}
\end{align}

In the classical limit $\hbar\to0$, the noise strength is $\gamma T_{eff}\sim\mathcal{O}(1)$
and the damping $\gamma\sim1/\sqrt{\hbar}\to\infty$, while the effective
temperature $T_{eff}\sim\sqrt{\hbar}\to0$. In this limit, the dissipation term leads to infinite damping, which originates from the feedback that we incorporated, and the system becomes completely static. Hence to get a nontrivial dynamics we keep $\hbar$ finite but small, as discussed in the main text. 

It is important to note that the effective temperature $T_{eff}$ solely arises from the measurement and feedback processes in Eq.\eqref{eq:DensityMatrixEvolution_S}, not due to any external thermal bath. Here we are considering an otherwise isolated system which  only interacts with the meters. The system could be initially prepared in a pure state and the time evolution [Eq.\eqref{eq:DensityMatrixEvolution_S}] preserves the purity of the many-body state of the system. In this case, for the semiclassical limit of small but nonzero $\hbar$, the system will approach a pure NESS in the longtime limit ($t_0\to-\infty$). However, remarkably, the effective distribution of the positions and momenta (classical components) of the oscillators in the pure NESS is described by a Boltzmann-Gibbs distribution $\sim \exp[-\mathcal{H}_s(\{x_i,p_i\})/T_{eff}]$, where the system Hamiltonian $\mathcal{H}_s$ is evaluated for classical phase-space configuration $\{x_i,p_i\}$.

The outcome of the measurements $\{\xi_i(t)\}$ drops out from the dynamics [Eq.\eqref{eq:LangevinEqn_S}] in the above semiclassical limit. In this limit, the measurement outcomes are 
\begin{align}
    \xi_i(t)=x_i(t),
\end{align}
i.e., pinned to their corresponding oscillator positions. This is because $\xi_i(t)$ is distributed according to $\exp{[-(1/2\Delta)\int dt(\xi_i-x_i)^2]}$ [see Eq.\eqref{eq:KeldyshAction_S}], with a mean $x_i(t)$ and variance $\Delta\sim \hbar^2$. As a result, in the semiclassical limit the fluctuations of the measurement outcome $|\xi_i(t)-x_i(t)|\sim \hbar$, and thus can be neglected up to $\mathcal{O}(1)$.  


\section{Gunsteren-Berendsen Method} \label{sec:GB_Method_S}
The dynamical equation in Eq.\eqref{eq:Dynamics_Gen} (main text), i.e.
\begin{align}
\ddot{x}_i+\gamma \dot{x}_i&=\frac{1}{m}\left[F_i(t)+\eta_i\right],
\end{align}
where $F_i(t)=-\partial V/\partial x_i$ is the force on the $i$-th particle, is simulated using a stochastic numerical scheme known as Gunsteren-Benrendsen (GB) method. We have followed the original paper \cite{GB1982} to implement the method. According to this, $x(t_n + \Delta t)$ is is obtained from $x(t_n)$, $\dot{F}(t_n)$ and $F(t_n)$ (omitting the index $i$ for brevity) as follows 
\begin{multline}
    x(t_n + \Delta t) = x(t_n)\Big[1 + \exp(-\gamma\Delta t) \Big] - x(t_n-\Delta t)\exp(-\gamma\Delta t)+ \frac{F(t_n) \Delta t}{m\gamma}\Big[ 1 - \exp(-\gamma\Delta t)\Big]\\ +\frac{\dot{F}(t_n) \Delta t}{m\gamma^2}\Bigg[\frac{1}{2}\gamma\Delta t \Big[1 + \exp(-\gamma\Delta t)\Big] - \Big[ 1 - \exp(-\gamma\Delta t)\Big]\Bigg] + X_n(\Delta t) + \exp(-\gamma\Delta t)X_n(-\Delta t) + \mathcal{O}\left[(\Delta t)^4 \right].
    \label{eq:GB_Gen}
\end{multline} 
The above reduces to Verlet molecular dynamics algorithm for $\gamma \rightarrow 0$ as
\begin{equation}
    x(t_n + \Delta t) = 2x(t_n) - x(t_n - \Delta t) + \frac{F(t_n) (\Delta t)^2}{m} +  \mathcal{O}\left[(\Delta t)^4 \right].
    \label{eq:GB_Verlet}
\end{equation} 
In the overdamped limit $\gamma \gg 1$ the algorithm reduces to \begin{equation}
    x(t_n + \Delta t) = x(t_n) + \frac{1}{m\gamma}\left[F(t_n)\Delta t + \dot{F}(t_n)\frac{1}{2}(\Delta t)^2 \right] + X_n(\Delta t).
    \label{eq:GB_Overdamped}
\end{equation} 
We can use above algorithm for smaller values $\gamma\Delta t$ with some subtlety as mentioned below. The velocity at each time step is calculated using positions at next and past instant of time as follows:
\begin{equation}
     v(t_n) = \Bigg[\Big[ x(t_n + \Delta t) - x(t_n - \Delta t)\Big] + \frac{F(t_n)}{m\gamma^2}G(\gamma\Delta t) - \frac{\dot{F}(t_n)}{m\gamma^3}G(\gamma\Delta t) + \Big[X_n(-\Delta t) - X_n(\Delta t) \Big]\Bigg]H(\gamma\Delta t)/\Delta t.
\end{equation} 
Here $X_n(\Delta t)$ and $X_n(-\Delta t)$ are the Gaussian random numbers with mean zero and widths that are function of coefficients $E, C, G$ and $H$, temperature and $\gamma$. The expressions for these coefficients can be found in ref.\onlinecite{GB1982}. For $\gamma\Delta t \ll 1$, due to presence of exponential terms, the numerical precision becomes an issue, and we need to use series expansion of coefficients (see appendix of ref.\onlinecite{GB1982}). While we have used the original coefficients to equilibrate the system at temperature $T$ and obtain the initial configurations with large $\gamma$, series expansion for the coefficients have been used to evolve the trajectories and calculate the cOTOC. 

\section{Thermal initial configurations}\label{sec:Thermal_IC}
Starting from an initial condition with all $x_i$'s and $p_i$'s zero, equilibrium configurations of of the system are generated in two ways. In one case, we used original GB scheme as in Eq. \eqref{eq:GB_Gen} to simulate Langevin dynamics of the model for moderate $\gamma$ and time step $\Delta t = 0.2$, without the series expansion of the coefficients. Equilibration is assumed when kinetic energy per particle reaches the equipartition value $E_{kin}/N = k_{B}T/2$. We generate $10^5$ initial configurations, separated by $100 \Delta t$ for independent uncorrelated results. 

We also use the overdamped limit of Langevin dynamics with a higher value of noise strength $\gamma \sim 20.0$ [Eq. \eqref{eq:GB_Overdamped}] with the time step $\Delta t=0.2$. Here too we do not need to use the series expansion of the coefficients that appear in the appear in the implementation of the GB algorithm \cite{GB1982}, as discussed in the preceding section. Equilibration was confirmed by checking that the potential energy per particle reaches a steady state. Thus we generate $10^5$ initial configurations of the positions of the particles. The momentum $p_i$ of each oscillator at equilibrium is drawn from a Gaussian distribution of width $\sqrt{k_B T/m}$, and these along with the position coordinates,  provides the thermal initial conditions $\{x_i(0), p_i(0) \}$ at temperature $T$. This way we generate the thermal configurations for different values of the parameters in the Hamiltonian, i.e., interaction $u$ for the nonintegrable chain, and $a$ and $b$ for the Toda chain. The procedure is repeated for four different system size $L = 1024, 800, 512$ and $256$. 

\begin{figure}[ht]
	\centering
	\includegraphics[width=0.6\linewidth]{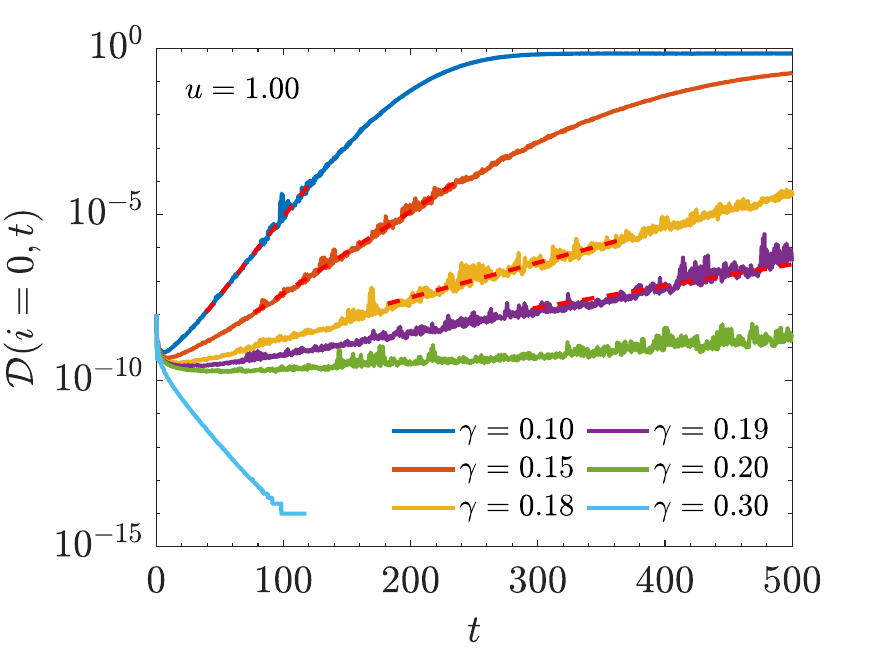}
		\caption{{\bf cOTOC across the noise-driven transition in the nonintegrable chain:} cOTOC at site $i=0$, $\mathcal{D}(i=0,t)$, is shown across the stochastic synchronization transition as a function of dissipation strength $\gamma$ for a fixed interaction strength  $u = 1.0$. The critical value for the transition is $\gamma_c \simeq  0.20$.} 
	\label{fig:Lyapunov_gamma_S}
\end{figure}

\begin{figure}[ht]
	\centering
	\includegraphics[width=0.6\linewidth]{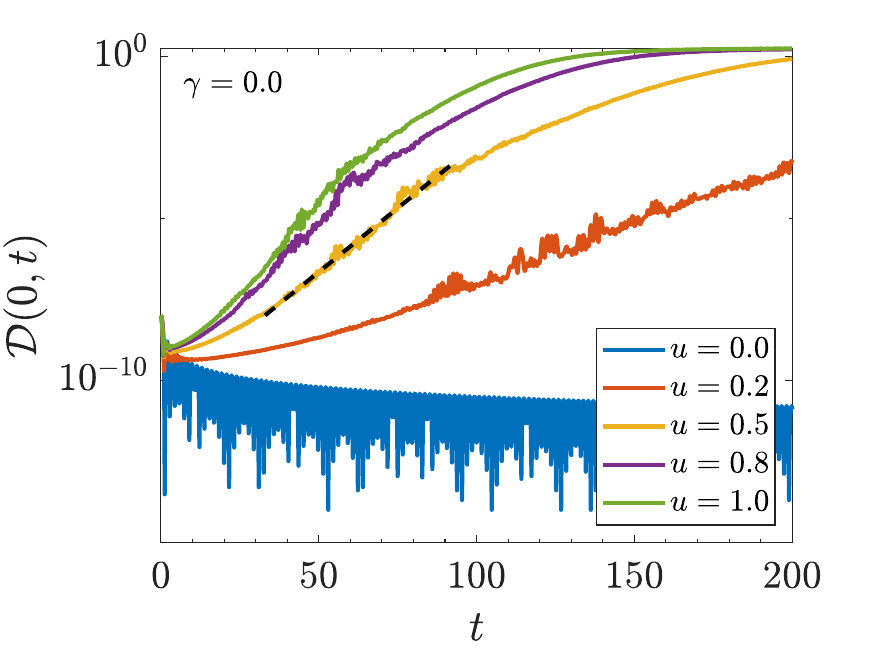}
		\caption{\textbf{cOTOC in the absence of noise and dissipation in the nonintegrable chain:} $\mathcal{D}(i=0,t)$ at $\gamma=0$ for different values of $u$, including the harmonic limit $u=0$. Black dashed line shows the fitted exponential regime. } 
	\label{fig:Lyapunov_gamma0_S}
\end{figure}

\section{Classical OTOC}\label{sec:Extraction_S}
The diagnostics of chaos in our calculation is captured by the classical OTOC (cOTOC) or the trajectory difference defined in Eq.\eqref{eq:Decorr_p}, main text. We define it in terms of momentum  to avoid the continuous drift in the center of mass of the chain under periodic boundary condition.  We start with an equlibrated configuration $\{x^\mr{A}_i(0), p^\mr{A}_i(0) \}$ and obtain the initial condition for the the other trajectory using $p^{B}_i(0) = p^{A}_i(0) + \delta_{i,0} \varepsilon$ with $\varepsilon = 10^{-4}$, i.e. the trajectory $B$ differs by a small amount from $A$ at only $i=0$ site. Each copy is evolved with the Langevin dynamics, and at each time and space points we calculate the trajectory difference in momentum coordinate as $\langle [p_i^{{A}}(t) - p_i^{{B}}(t)]^2 \rangle$, where we have used $10^5$ initial configurations. Initially momenta of all oscillators are exactly the same in two copies except at the perturbing site $i = 0$, and we get $\mathcal{D}(i, 0) = \varepsilon^2\delta_{i,0}$, i.e., zero at all sites except $i = 0$.

 \begin{figure}[ht]
	\centering
	\includegraphics[width=0.6\linewidth]{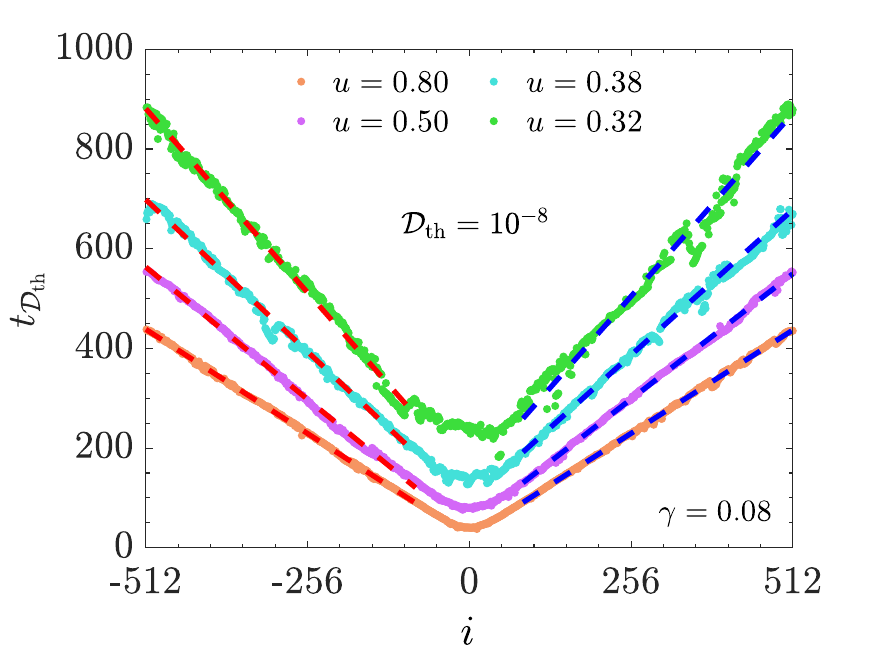}
		\caption{\textbf{Extraction of butterfly velocity:} Light cones for the threshold value $\mathcal{D}_{\mathrm{th}} = 10^{-8}$ at different interaction strengths $u>u_c$ for a fixed noise strength $\gamma = 0.08$. These contours are fitted (dashed lines) with $t_{\mathcal{D}_{\mathrm{th}}} = t_0 \pm i/v^{\mr{L(R)}}_{B}$ on the left (right) side of $i=0$ and the butterfly velocity is extracted from $v_{B} = (v_{B}^{\mr{L}} + v_{B}^{\mr{R}})/2$.} 
	\label{fig:ButtVeolo_Extract_S}
\end{figure}

 \begin{figure}[ht]
	\centering
	\includegraphics[width=0.6\linewidth]{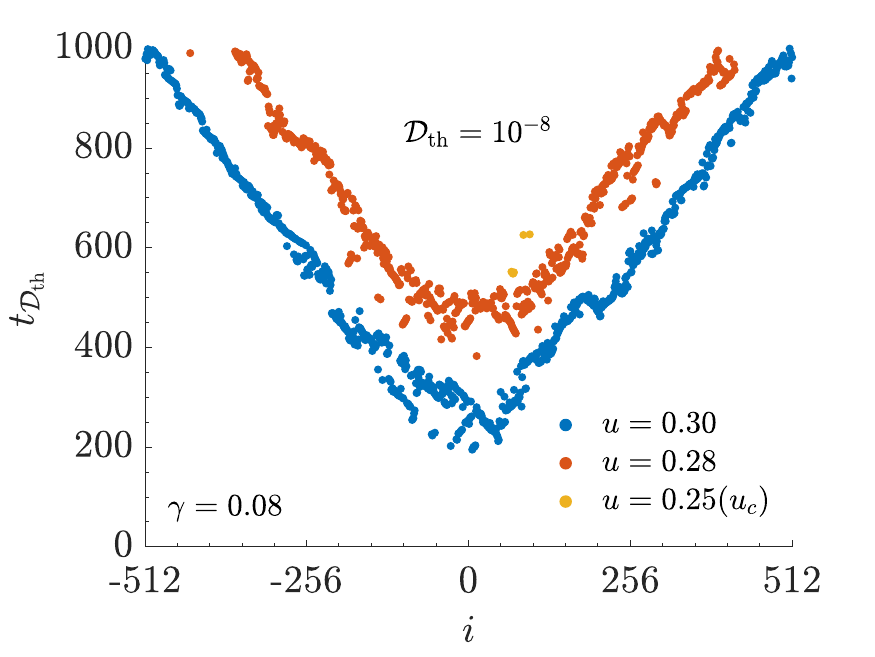}
		\caption{\textbf{Threshold light cones close to the transition:} The light cones becomes more and more distorted as we approach the transition and hence it is difficult to extract $v_B$ close to $u_c$.} 
	\label{fig:LightCones_uc_S}
\end{figure}

\begin{figure}[ht]
	\centering
	\includegraphics[width=0.7\linewidth]{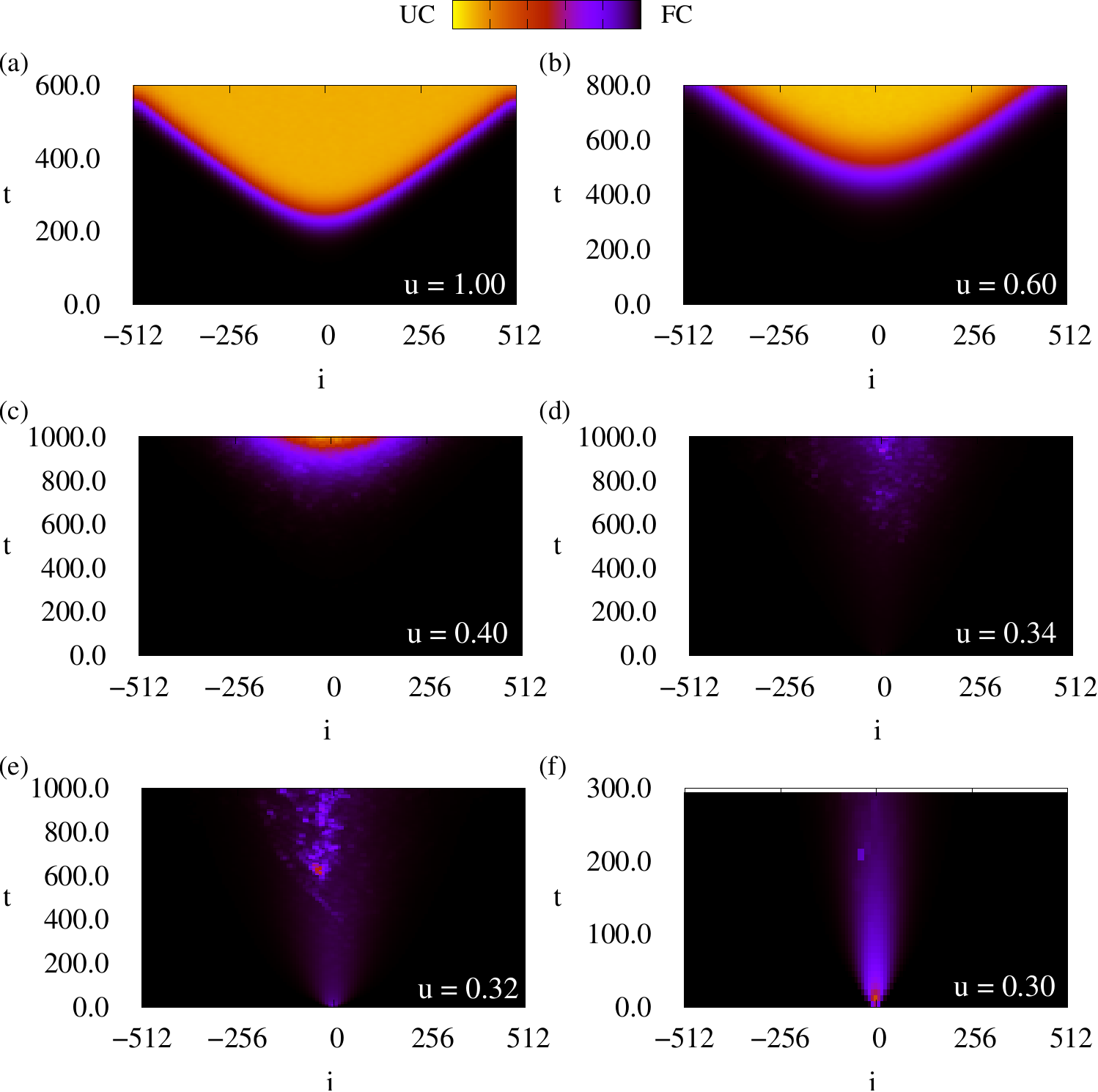}
		\caption{\textbf{Transition in the light cone for the nonintegrable chain:} Chaotic to nonchaotic transition in terms of light cone for cOTOC (color). Above plots are shown with $u$ varied and the noise strength $\gamma = 0.10$ fixed for system size $L=1024$. Critical value of interaction is $u_c = 0.32$, below which the transition to the nonchaotic phase takes place.}
	\label{fig:LC_Transtion_g0.10}
\end{figure}

To extract Lyapunov exponent, we fit the trajectory difference at the perturbing site with an exponential form, i.e., $\mathcal{D}(i = 0, t) \sim e^{2\lambda_Lt}$. If the model is in the chaotic regime, this exponential form would fit the perturbation growth correctly within an intermediate time window and give a positive Lyapunov exponent. This fitting procedure is repeated for different parameter values in the nonintegrable chain to extract $\lambda_L$ across the transitions with respect to $u$ for $\gamma\neq 0$ [Fig.\ref{fig:Panel1_Schematics}(c), main text], and as function of $\gamma$ (Fig.\ref{fig:Lyapunov_gamma_S}) for $u\neq0$. For comparison, we also plot $\mathcal{D}(i=0,t)$ for $\gamma=0$ in Fig.\ref{fig:Lyapunov_gamma0_S}. In this case the cOTOC grows exponentially with $\lambda_L>0$ for any finite $u$, and decays only for the harmonic chain $u=0$.

To extract the butterfly velocity in the chaotic phase, we plot the chaos onset contour by finding the time required, $t_{\mathcal{D}_{\mr{th}}}(i)$, for $\mathcal{D}(i, t)$ at each site to reach a threshold value $\mathcal{D}_{\mr{th}} = 10^{-8} (= \varepsilon^2)$. For a ballistic spreading of chaos front, $t_{\mathcal{D}_{\mr{th}}}$ shows a linear dependence on site index $i$ (Fig.\ref{fig:ButtVeolo_Extract_S}), i.e. a light cone exists. Hence, we fit the contour with a ballistic form $t_{\mathcal{D}_{\mr{th}}} = t_0 + i/v^{\mr{L(R)}}_B$ where $v^{\mr{L(R)}}_B$ is the butterfly velocity extracted from the left (right) side of $i=0$. Here $t_0$ is an onset time \cite{Ruidas2021} that the cOTOC takes to reach the threshold value at $i=0$, i,e., $\mathcal{D}(0, t_0) = \varepsilon^2$. The velocity is obtained from $v_{B} = (v_B^{\mr{L}} + v_B^{\mr{R}})/2$ (Fig.\ref{fig:ButtVeolo_Extract_S}). As earlier, this process is repeated for different parameter values to extract $v_B$ across the transitions. The ballistic light cone progressively deteriorates approaching the transition (Fig.\ref{fig:LightCones_uc_S}) and completely vanishes once we reach the nonchaotic phase for $u < u_c$ (Fig.\ref{fig:LC_Transtion_g0.10}). We find similar transition in the light cones for $\gamma > \gamma_c$.

\begin{figure}[ht]
	\centering
	\includegraphics[width=0.6\linewidth]{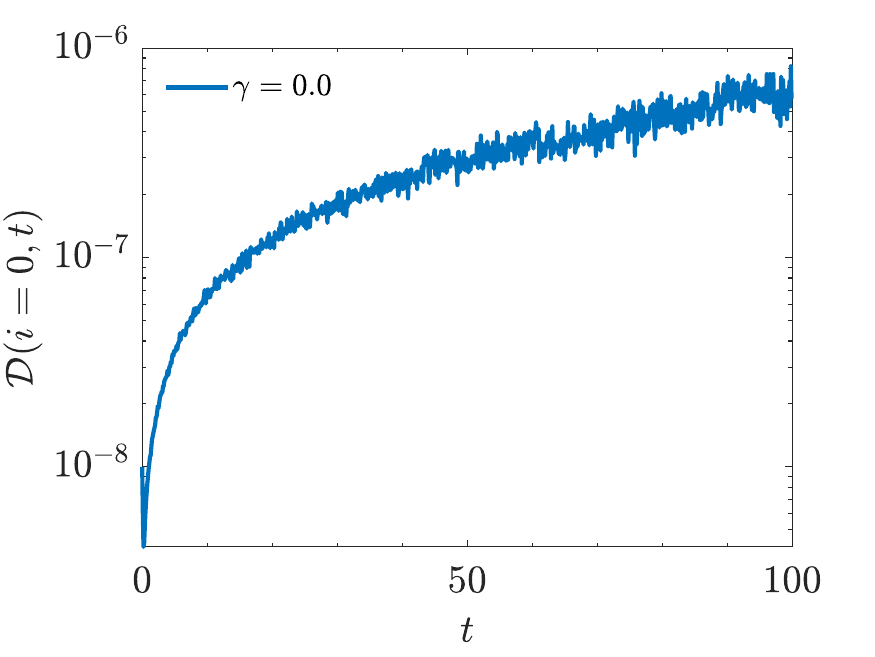}
		\caption{\textbf{cOTOC in the integrable Toda chain}. $\mathcal{D}(0,t)$ for $\gamma=0$ for the Toda chain with $a=0.07$ and $b=15$.} 
	\label{fig:Toda_D0Evol_gamma0_S}
\end{figure}

\begin{figure}[ht]
	\centering
	\includegraphics[width=0.6\linewidth]{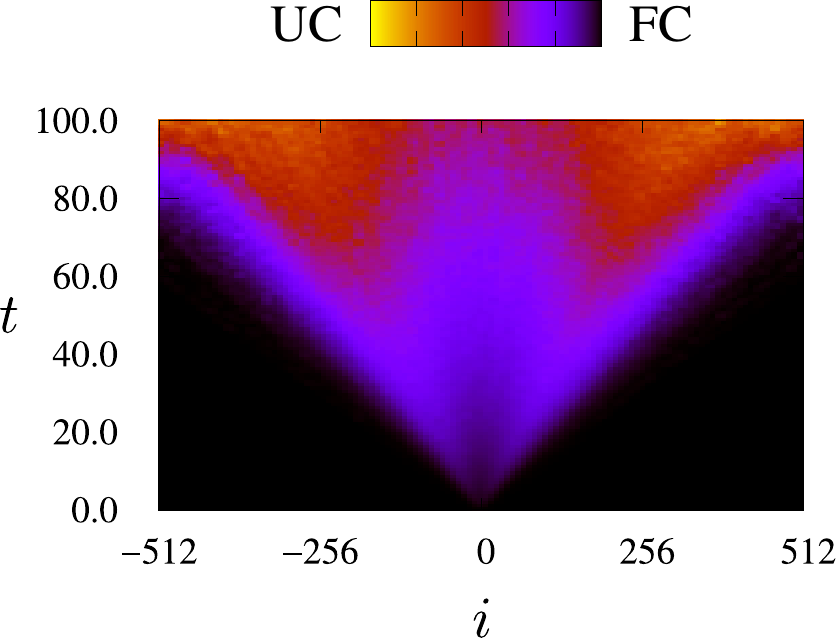}
		\caption{\textbf{Lightcone in the integrable Toda chain}. cOTOC $\mathcal{D}(i,t)$ (color) for for $\gamma=0$ for the Toda chain with $a=0.07$ and $b=15$.} 
	\label{fig:Toda_Lightcone_gamma0_S}
\end{figure}

\begin{figure}[ht]
	\centering
	\includegraphics[width=0.6\linewidth]{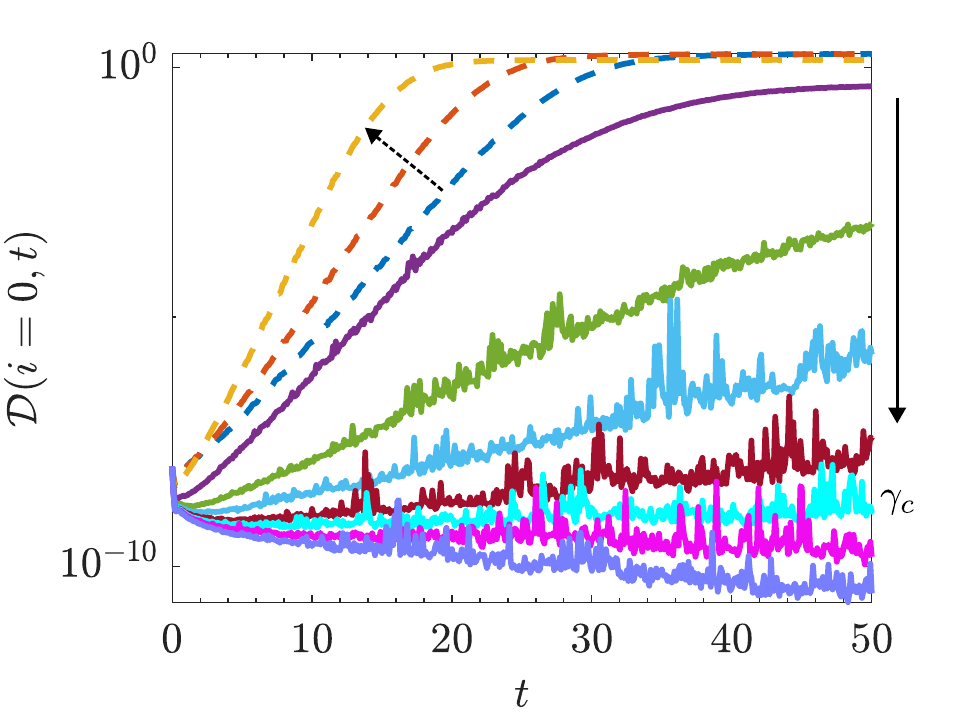}
		\caption{\textbf{cOTOC across the noise-driven transition in the Toda chain:} Time evolution of cOTOC at $i=0$ for Toda model as a function of dissipation strength $\gamma = 0.02, 0.05, 0.50 (=\gamma_m), 2.00, 2.70, 3.00, 3.20, 3.30 (=\gamma_c), 3.40, 3.50$. The dashed line plots are for $\gamma < \gamma_m$ when the chaos growth rate increases with $\gamma$ before reaching a maximum at $\gamma_m$. After that (solid lines) $\lambda_L$ decreases approaching the transition at $\gamma_c = 3.30$ (cyan). Arrows specify the direction of increasing $\gamma$ in the plot. } 
	\label{fig:Toda_D0Evol_Transition_S}
\end{figure}

In a similar way, we extract $\lambda_L$ and $v_B$ for the integrable model of the Toda chain. For $\gamma=0$, the trajectory difference does not show any exponential growth, indicating a nonchaotic phase (Fig.\ref{fig:Toda_D0Evol_gamma0_S}) with $\lambda_{\mathrm{L}} = 0$. However, the initial perturbation still spreads ballistically with a nonzero velocity $v_{B}$, extracted in a similar fashion as earlier, as shown in the Fig.\ref{fig:Toda_Lightcone_gamma0_S}.  Interesting features show up when noise is added to the dynamics. As it can be seen from Fig.\ref{fig:Toda_D0Evol_Transition_S}, the dynamics becomes chaotic, showing an exponential growth of cOTOC, as soon as any finite noise ($\gamma\neq 0$) is added. The growth rate increases rapidly as we increase the noise strength and at a value $\gamma_m$, the Lyapunov exponent attains a maximum value [see Fig.\ref{fig:Panle4_TodaResults}(a) in the main text]. After this point, $\lambda_L$ starts to decrease. At a critical value of noise strength $\gamma_c$, the model undergoes the synchronization transition to a nonchaotic phase. This transition is very similar to the one that we witnessed in the nonintegrable chain, i.e., we see a continuous decrease of $\lambda_L$ from positive to negative values. We extract $v_B$ across the transition as earlier. Unlike the nonintegrable model, here $v_B$ shows a sharp jump from its value at $\gamma=0$ for any non zero noise strength [see Fig.\ref{fig:Panle4_TodaResults}(c), main text] indicating a singular behaviour as $\gamma\to0^+$. This jump is then followed by a monotonic decrease approaching $\gamma_c$ [see Fig.\ref{fig:Panle4_TodaResults} (c), main text].

\section{Finite size scaling of butterfly velocity} \label{sec:Scaling_S}

As shown in the main text Fig.\ref{fig:Panel3_ButtVelo}(b) (inset), the butterfly velocity exhibits finite size dependence near the transition. We perform a finite-size scaling analysis with system size for $v_B$, as shown in the main text Fig.\ref{fig:Panel3_ButtVelo}(b) (main panel). Here we discuss the details of the scaling procedure. We write $v_B$ as 
\begin{equation}
    v_B(u, L) = L^{-\Delta_v}\mathcal{F} [\delta u L^{1/\nu}]
    \label{eq:vB_Scaling_S}
\end{equation} 
where $\mathcal{F}(x)$ is a universal scaling function and $\delta u=u-u_c>0$. The above can be written as $v_B(u,L)=L^{-\Delta_v}\mathcal{F}'(L/\xi)$ in terms of another scaling function $\mathcal{F}'(x)$ with a correlation length $\xi(\delta u)\sim (\delta u)^{-\nu}$. We find the optimized exponents $\nu$ and $\Delta_v$ so that the data for $v_B$ for all $u>u_c$ and $L$ collapses into a single curve described by Eq.\eqref{eq:vB_Scaling_S}. This is done by minimizing the $\chi^2$-function in multi-parameter space, defined as
\begin{equation}
    \chi^2(\{ c_k\}; \{d_k\}; \Delta_v) = \sum_{L, u} \Big[v_B(u, L) - L^{-\Delta_v}\mathcal{F}'(L/\xi(\delta u)) \Big]^2,
    \label{eq:ChiSq_S}
\end{equation} 
where $v_B(u, L)$ are the data points at different values of $u$  and $L$. We assume a second-order polynomial for the scaling function $\mathcal{F}'(x \equiv L/\xi) = c_0 + c_1 x + c_2 x^2$ and  a third-order one for the correlation length $\xi(\delta u) = d_0 + d_1(\delta u) + d_2(\delta u)^2 + d_3(\delta u)^3$. For a fixed value of $u_c$, starting from an initial guess for the parameters, we can minimize the $\chi^2$ in  Eq.\eqref{eq:ChiSq_S} and obtain the optimized $c_k$'s, $d_k$'s and $\Delta_v$. The exponent $\Delta_v$ is obtained directly from this procedure. The correlation length exponent $\nu$ is extracted from the fit $\xi(\delta u) \sim (\delta u)^{-\nu}$ with a power law. We have checked that extracted critical exponents are robust with respect to a range of initial guesses for $\{c_k,d_k\}$. 

\begin{figure}[ht]
	\centering
	\includegraphics[width=0.6\linewidth]{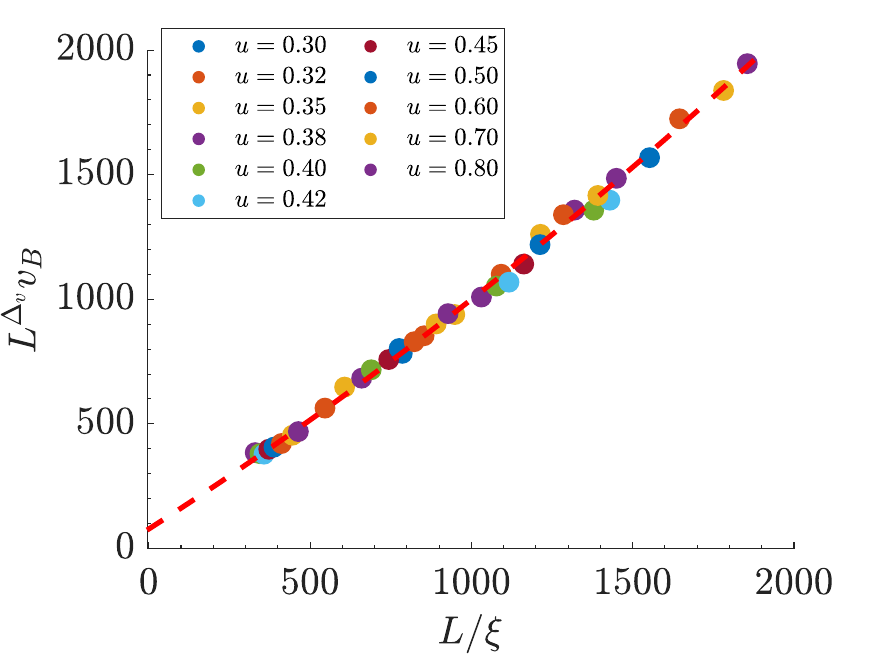}
		\caption{\textbf{Scaling collapse for butterfly velocity for $\gamma=0.08$ in the nonintegrable chain:} Scaling collapse of $L^{\Delta_v} v_{B}$ with $L/\xi$ for $u_c = 0.242$. The polynomial scaling function $\mathcal{F}'(x=L/\xi)$ is shown by red dashed line. } 
	\label{fig:FPoly_Optimized_S}
\end{figure}

\begin{figure}[ht]
	\centering
	\includegraphics[width=0.6\linewidth]{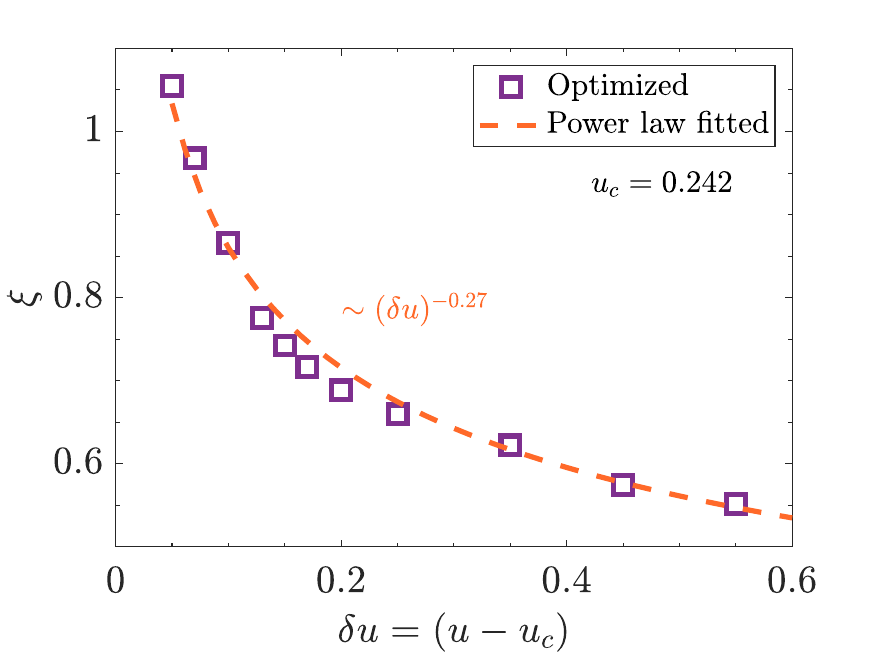}
		\caption{\textbf{Optimized correlation length and the extraction of $\nu$ from scaling collapse for $\gamma=0.08$ in the nonintegrable chain: } The correlation length $\xi$ extracted from the finite-size scaling is plotted as a function of $\delta u$ with $u_c = 0.242$. The fit of the data with a power law, i.e., $\xi \sim (\delta u)^{-\nu}$, gives the critical exponent $\nu \simeq 0.27$.} 
	\label{fig:Xi_Optimized_S}
\end{figure}

We have shown the results the scaling collapse for $\gamma = 0.08$ in the main text [Fig. \ref{fig:Panel3_ButtVelo}(b)(main panel)]. A typical collapse of $L^{\Delta_v}v_B$ with $L/\xi$ with the scaling function $\mathcal{F}'(x)$ is shown in Fig.\ref{fig:FPoly_Optimized_S}. The correlation length and associated exponent $\nu$ extracted form the scaling collapse are shown in Fig.\ref{fig:Xi_Optimized_S}. For the results shown here, and in Fig. \ref{fig:Panel3_ButtVelo}(b), main text, we have taken $u_c = 0.242$ which is close to the $u_c\simeq 0.25$ obtained from Lyapunov exponent results [Fig.\ref{fig:Panel2_transitions}(c), main text]. 

The quality of the data collapse [main text, Fig.\ref{fig:Panel3_ButtVelo}(b) (main panel)] remains reasonably good for a range of $u_c$ values between 0.25 to 0.21 as shown by $\chi^2$ values in Table \ref{tab:scaling}. As it can be seen from the table, $\nu$ varies over a range $\sim 0.25-0.35$ and $\Delta_v$ over $\sim 1.00-1.06$ for $u_c$ within the range 0.25-0.21. In this range, the $\chi^2$ value stays reasonably low $\sim 0.025$, which ensures validity of the scaling collapse. 

\begin{center}
\begin{table}
\renewcommand{\arraystretch}{1.5}
\begin{tabular}{ |c|c|c|c| }
 \hline
 $u_c$ & $\chi^2_{\mr{min}}$ & $\nu$ & $\Delta_v$ \\ 
 \hline\hline
 0.250 & 0.026 & 0.253 & 1.048\\ 
 \hline
 0.242 & 0.026 & 0.265 & 1.063\\ 
 \hline
 0.220 & 0.025 & 0.332 & 1.010\\
 \hline
 0.210 & 0.026 & 0.340 & 1.050\\
 \hline
 0.200 & 0.031 & 0.382 & 1.039\\
 \hline
\end{tabular}
\caption{{\bf Table of $\chi^2$ values and critical exponents for different choices of $u_c$ in the finite-size scaling for the nonintegrable chain with $\gamma=0.08$.}}\label{tab:scaling}
\end{table}
\end{center}

\section{Mean square displacement}
\begin{figure}[ht]
	\centering
	\includegraphics[width=0.6\linewidth]{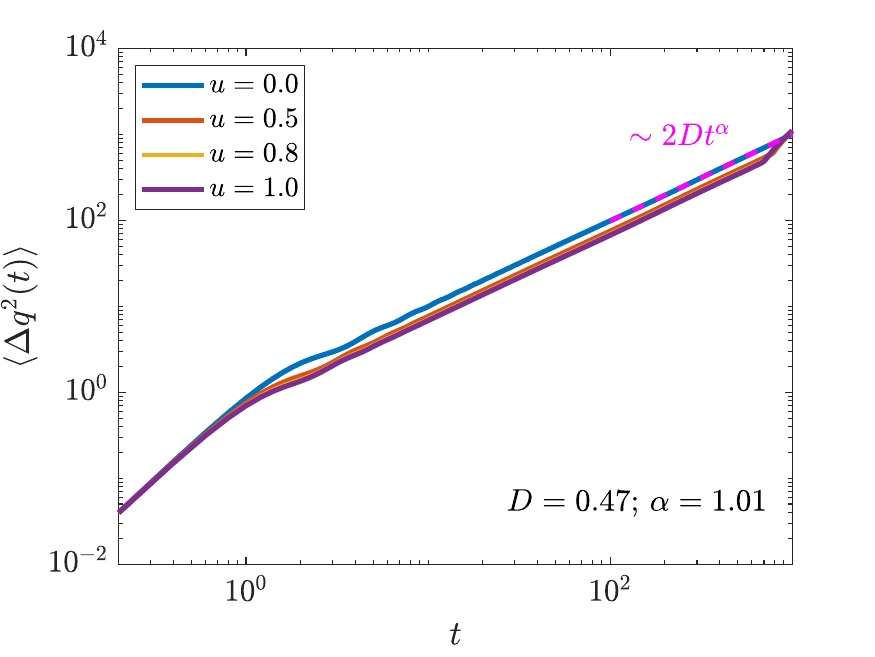}
	\caption{{\bf Mean square displacement in the nonintegrable chain in the absence of noise/dissipation:} $\langle \Delta q^2(t)\rangle$ as a function of $t$ for different interaction $u$'s at $\gamma=0$. The MSD is well fitted for all values of $u$, with a diffusive behaviour $\sim 2Dt^\alpha$ with $\alpha=1$ and a diffusion constant $D \sim 0.5$, expected from ref.\onlinecite{Lee1985}.}
	\label{fig:MSD_gamma0_S}
\end{figure} 

To verify that the usual dynamical quantities calculated from a single trajectory does not show any signature of the dynamical transition seen in the cOTOC, we calculate the mean square displacement (MSD),
\begin{equation}
    \langle \Delta q^2(t)\rangle  = \frac{1}{N}\sum_{j = 1}^{N}\left\langle\big[x_j(t) - x_j(0) \big]^2\right\rangle,
    \label{eq:MSD}
\end{equation} 
for a single trajectory generated from the initial thermal configurations. Here we use the series expansions for the coefficients in the GB algorithm\cite{GB1982} (see Sec.\ref{sec:GB_Method_S}) and a time step $\Delta t = 0.05$. We show the results for $\gamma = 0$ in Fig.\ref{fig:MSD_gamma0_S}. As shown in ref.\onlinecite{Lee1985}, the harmonic chain $u=0$ with periodic boundary condition exhibits a surprising diffusive behaviour $\langle \Delta q^2(t)\rangle\sim t$ with a diffusion constant $(T/2m\omega_0)$, as we verify. Here $\omega_0=\sqrt{\kappa/m}$ and $\kappa$ the harmonic coupling between the oscillators. The diffusive behaviour persists for $u\neq 0$ with little change of the diffusion constant, as shown in Fig.\ref{fig:MSD_gamma0_S}. 

\begin{figure}[ht]
	\centering
	\includegraphics[width=0.6\linewidth]{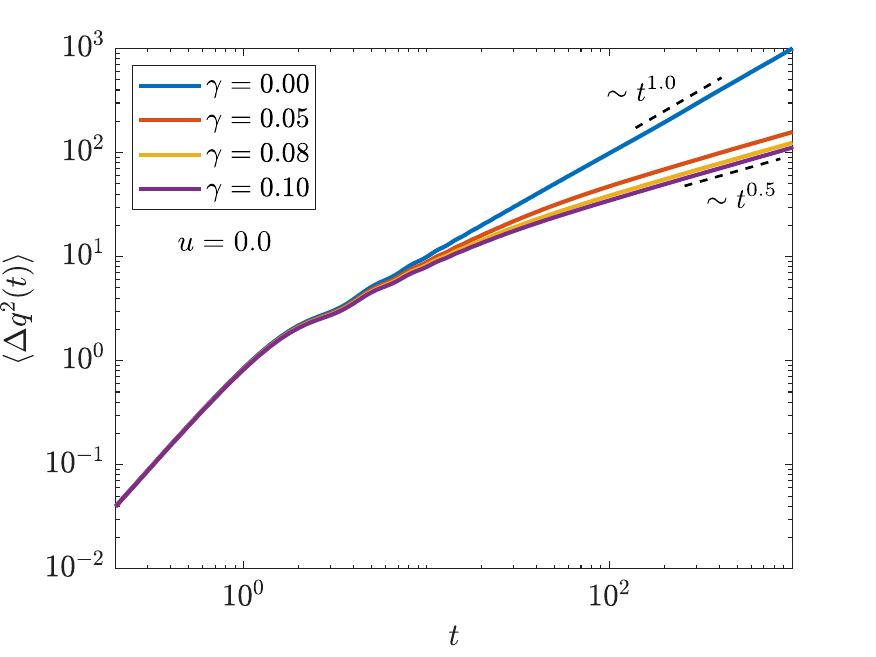}
	\caption{{\bf Mean square displacement in the harmonic chain in the presence of noise/dissipation:} $\langle \Delta q^2(t)\rangle$ as a function of $t$ for different $\gamma$'s for $u=0$. The MSD is well fitted with a subdiffusive form $\sim \sqrt{t}$ for all values of $\gamma$, except $\gamma=0$.}
	\label{fig:MSD_u0_S}
\end{figure}

\begin{figure}[ht]
	\centering
	\includegraphics[width=0.6\linewidth]{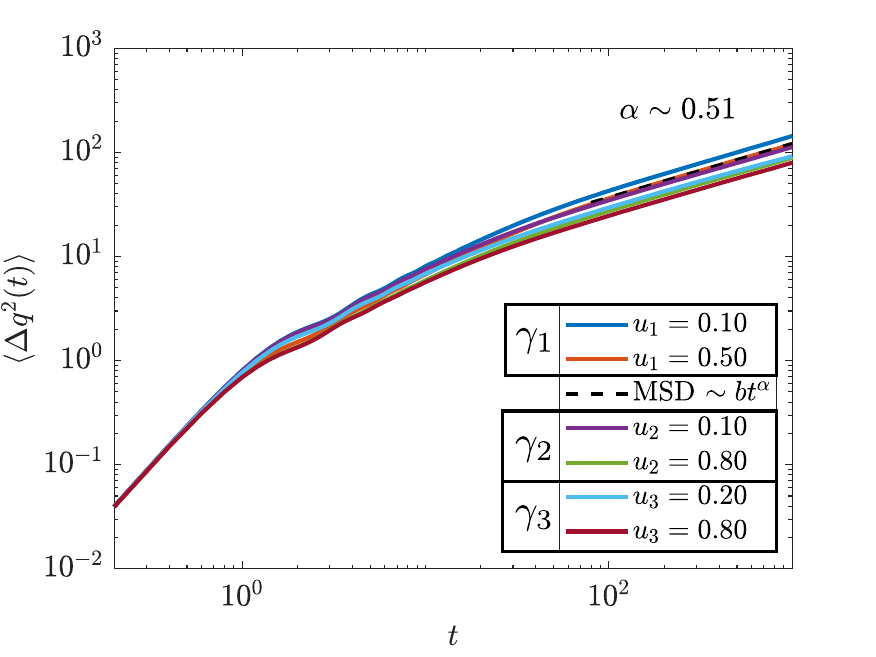}
	\caption{{\bf Mean square displacement in the nonintegrable chain in the presence of noise/dissipation:} $\langle \Delta q^2(t)\rangle$ as a function of $t$ for different $u$'s for three different values of $\gamma$; $u_1$, $u_2$ and $u_3$ denote chosen values of interaction strength for $\gamma = 0.05$, 0.08 and 0.10, respectively. In all cases, The MSD is well fitted with a subdiffusive form $\sim \sqrt{t}$.}
	\label{fig:MSD_u_S}
\end{figure}

Switching on a nonzero dissipation immediately changes the time-dependence of MSD to a subdiffusive behaviour $\langle \Delta q^2(t)\rangle\sim \sqrt{t}$, as shown in Fig.\ref{fig:MSD_u0_S} for the harmonic limit $u=0$. The subdiffusive behaviour can be understood from monomer subdiffusion in polymer dynamics \cite{Andrew2010}. The same subdiffusive behaviour remains after turning on interaction $u\neq 0$, as shown in Fig.\ref{fig:MSD_u_S} for three different values of $\gamma$.

\end{document}